\documentclass[lettersize,journal]{IEEEtran}
\usepackage{amsmath,amsfonts}
\usepackage{algorithm}
\usepackage{algpseudocode}
\usepackage{array}
\usepackage[caption=false,font=normalsize]{subfig}
\usepackage{textcomp}
\usepackage{stfloats}
\usepackage{url}
\usepackage{verbatim}
\usepackage{graphicx}
\usepackage{cite}
\usepackage{bm}
\usepackage{amssymb}
\usepackage[detect-weight,retain-explicit-plus]{siunitx} 
\usepackage{xcolor}
\usepackage{tikz}

\usepackage[absolute,overlay]{textpos}

\newcommand{\T}{^{\mathrm T}}

\newcommand{\E}[1]{\mathcal{E} \lbrace {#1} \rbrace}

\newcommand{\dg}[1]{\mathrm{diag}\left( {#1} \right)}
\newcommand{\x}{{\mathrm x}}
\newcommand{\s}{{\mathrm s}}
\newcommand{\n}{{\mathrm n}}

\newcommand{\rev}[1]{{\textcolor{black}{#1}}}
\newcommand{\revv}[1]{{\textcolor{black}{#1}}}

\begin{document}

\begin{textblock*}{19cm}(1cm,0.5cm) 
\small{\copyright~2023 IEEE. Personal use of this material is permitted. Permission from IEEE must be obtained for all other uses, in any current or future media, including reprinting/republishing this material for advertising or promotional purposes, creating new collective works, for resale or redistribution to servers or lists, or reuse of any copyrighted component of this work in other works.}
\begin{center}
    DOI 10.1109/TASLP.2023.3240657
\end{center}
\end{textblock*}

\title{Direct and Residual Subspace Decomposition of Spatial Room Impulse Responses}

\author{Thomas~Deppisch,~\IEEEmembership{Graduate Student Member,~IEEE},
        Sebastià~V.~Amengual~Garí,
        Paul~Calamia,
        and~Jens~Ahrens,~\IEEEmembership{Senior Member,~IEEE}
\thanks{This research was supported by Reality Labs Research at Meta.}
\thanks{Thomas~Deppisch and Jens~Ahrens are with the Chalmers University of Technology, 412 96 Gothenburg, Sweden (e-mail: thomas.deppisch@chalmers.se; jens.ahrens@chalmers.se).}
\thanks{Sebastià~V.~Amengual~Garí and Paul~Calamia are with Reality Labs Research, Meta, Redmond, WA 98052, USA (e-mail: samengual@meta.com; pcalamia@meta.com).}
\thanks{Manuscript received May 09, 2022; revised XXXX XX, 2023.}}


\IEEEpubid{0000--0000/00\$00.00~\copyright~2023 IEEE}

\maketitle

\begin{abstract} 
Psychoacoustic experiments have shown that directional properties of the direct sound, salient reflections, and the late reverberation of an acoustic room response can have a distinct influence on the auditory perception of a given room. Spatial room impulse responses (SRIRs) capture those properties and thus are used for direction-dependent room acoustic analysis and virtual acoustic rendering. This work proposes a subspace method that decomposes SRIRs into a direct part, which comprises the direct sound and the salient reflections, and a residual, to facilitate enhanced analysis and rendering methods by providing individual access to these components. The proposed method is based on the generalized singular value decomposition and interprets the residual as noise that is to be separated from the other components of the reverberation. Large generalized singular values are attributed to the direct part, which is then obtained as a low-rank approximation of the SRIR. By advancing from the end of the SRIR toward the beginning while iteratively updating the residual estimate, the method adapts to spatio-temporal variations of the residual. The method is evaluated using a spatio-spectral error measure and simulated SRIRs of different rooms, microphone arrays, and ratios of direct sound to residual energy. The proposed method creates lower errors than existing approaches in all tested scenarios, including a scenario with two simultaneous reflections. A case study with measured SRIRs shows the applicability of the method under real-world acoustic conditions. A reference implementation is provided.
\end{abstract}


\section{Introduction}
\IEEEPARstart{D}{irectional} properties of acoustic environments have been subject to extensive research in recent years as they are a key factor in human auditory perception. It was found that salient reflections, i.e., reflections with sufficiently high energy, can have an individual impact on the perceived spatial impression~\cite{Barron1971}, the apparent source position and width~\cite{Olive1989}, and the timbre~\cite{Bech1995}. On the other hand, statistical properties of the reverberation, such as the overall energy, its angular distribution, the direct-to-reverberant energy ratio, and the reverberation time, influence the perceived envelopment~\cite{Bradley1995}, source distance~\cite{Zahorik2002}, and room size~\cite{Hameed2004, Yadav2013}. Auditory perception is also influenced by the directional energy decay of late reverberation~\cite{Romblom2016,Alary2021}. 

Spatial room impulse responses (SRIRs) capture the directional properties of an acoustic environment and facilitate the analysis and reproduction thereof. Note that the term SRIR is heavily used in literature but is defined inconsistently. We refer to an SRIR as a set of room impulse responses that is captured by a single microphone array to facilitate the directional analysis or auralization for a single source position and from the perspective of a single receiver position. Hence, suitable microphone arrays have a small aperture, typically smaller than \SI{0.5}{m}. No other requirements are imposed on the array geometry; however, due to their wide commercial availability and the possibility to perform a spherical harmonic (SH) decomposition of the array signals~\cite{Rafaely2019}, often spherical microphone arrays are used.

Motivated by their perceptual relevance, SRIR-based directional room acoustic analysis and virtual acoustic rendering methods specifically target salient reflections and directional statistic properties of the reverberation. Common analysis objectives include the direction-of-arrival (DOA) estimation of reflections~\cite{Khaykin2012, Huleihel2013, Tervo2015} and the directional energy decay of the reverberation~\cite{Gover2004,Berzborn2018, Patynen2013, Alary2021}.

SRIR-based virtual acoustic renderers reproduce the acoustics of an environment by convolving a processed SRIR with source signals. They target multi-channel loudspeaker and/or binaural headphone playback. The renderers either analyze the direction of the frequency-dependent instantaneous acoustic intensity~\cite{Merimaa2005} or use a broadband DOA estimator~\cite{Tervo2013, Zaunschirm2018a, AmengualGari2020} to impose spatial information onto an omnidirectional signal. An extension of~\cite{Merimaa2005} generalizes the method using higher-order SHs and processing in angular sectors~\cite{McCormack2020}. A recently proposed method~\cite{Puomio2021a} analyzes DOAs of reflections as in~\cite{Tervo2013} but explicitly cuts out salient reflections from the omnidirectional RIR to resynthesize the early part of an SRIR. The methods render diffuse reverberation either implicitly by a fast modulation of the reproduction direction, or explicitly by using a diffuseness estimate and decorrelating diffuse signal parts to create multi-channel loudspeaker signals. Instead of the decorrelation, the late reverberation might also be replaced by filtered noise~\cite{Meyer-Kahlen2021}. Other methods analyze SRIRs to generate parametric synthetic reverberation~\cite{Coleman2017, Stade2017, Brinkmann2020}.
\IEEEpubidadjcol

While most of the renderers are designed to accurately reproduce salient reflections and diffuse reverberation, none of the existing methods achieves an explicit separation of salient reflections from the SRIR while preserving the spatio-temporal properties of both the reflections and the residual. To overcome this limitation, the authors recently proposed to use the spatial subtraction method to subtract salient reflections from an SRIR by using a beamformer and a plane-wave prototype~\cite{Deppisch2021b}. \rev{The spatial subtraction method was initially proposed for the separation of direct and diffuse parts in sound scenes~\cite{Politis2018} and was extended in~\cite{Deppisch2021b} by employing a comprehensive microphone-array signal model that includes the impacts of scattering and spatial aliasing, which improved the separation performance when applied to SRIRs.}

The current work proposes a subspace method to separate SRIRs into a direct part, comprising the direct sound and salient reflections, and a residual. \rev{The proposed method is shown to improve the separation performance in comparison to the spatial subtraction method as it avoids the error-prone estimation of reflection parameters. It is free from typical assumptions, such as reflections being plane waves and late reverberation being isotropic, and does not rely on parameter estimation regarding, e.g., the number of simultaneous reflections and their DOAs.} In consequence, the method also does not provide an estimation of such parameters but rather generates two SRIRs, one containing the direct part and the other containing the residual, that can then be analyzed and processed independently. \rev{Nevertheless, the method may improve the performance of existing parameter estimation algorithms by applying it as pre-preprocessing.}

\rev{By providing an explicit separation of the direct part and the residual, the method facilitates advanced rendering and extrapolation strategies of SRIRs. A perceptual pilot study of such extrapolation strategies suggests that an efficient SRIR extrapolation to different positions in a room may be achieved using the proposed method by combining a residual SRIR from a single measurement with salient reflections from the target position~\cite{Deppisch2022b}.}

\section{Subspace Decomposition Theory}\label{sec:subspace_theory}
\subsection{General Principle}
Subspace methods in array processing are based on the assumption that target signals only occupy a limited subspace of the full signal space that is spanned by the multiple, noisy sensor readings. The methods reduce noise by confining the noisy signal to a subspace containing a superposition of signal and noise, called the \emph{signal subspace}, while disregarding components in the orthogonal \emph{noise subspace} that are solely attributed to the noise~\cite{Schmidt1986,Roy1989a}. 

Subspace methods essentially exploit the Eckart-Young-Mirsky theorem~\cite{Eckart1936} to find the best low-rank approximation of a signal matrix. In array signal processing, this was first applied by Tufts et al.~\cite{Tufts1982} and they showed that the low-rank approximation can either be performed via the singular value decomposition (SVD) of the data matrix or by an orthogonal projection using eigenvectors of the covariance matrix.
Later, the principle was exploited in beamforming~\cite[Ch.~6.8]{VanTrees2002} and parameter estimation~\cite{Schmidt1986,Roy1989a}. It was also applied in speech enhancement, first in single-channel~\cite{Dendrinos1991, Ephraim1995} and later in array-based methods~\cite{Asano2000,Doclo2002b}. 
In speech enhancement, noise components in the signal subspace are typically further reduced using a signal-dependent post-filter and several estimators have been proposed for that purpose~\cite{Hermus2007}. Those estimators reduce noise at the cost of increased signal distortion and thus will be disregarded in this work.

To motivate subspace methods mathematically, it is beneficial to analyze the covariance matrix of a noisy array signal. Let $\bm x(t)$ be a length-$M$ vector containing the signals that are captured by $M$ microphones \rev{at the discrete time $t$}. Following a convolutive multiple-input-multiple-output (MIMO) signal model~\cite[Ch.~2.1.4]{Huang2006}, the array signals are convolutive mixtures of the source signals $\bm s(t)$ plus additive noise $\bm n(t)$,
\begin{equation}\label{eq:MIMO_sig_model}
    \bm x(t) = \bm H \bm s(t) + \bm n(t) \, .
\end{equation}
In each row, the $M \times QN$ matrix $\bm H$ contains $Q$ finite impulse response (FIR) filters of length $N$ that describe the transfer paths from each of the $Q$ sources to one of the $M$ microphones. Accordingly, $N$ observations of each source signal are stacked in the length-$QN$ source signal vector $\bm s(t)$ such that $\bm H \bm s(t)$ describes the convolution of the source signals with the FIR filters at the time $t$. The noise vector $\bm n(t)$ is of length $M$ and contains one noise observation per microphone \rev{at the time $t$}. \rev{This general signal model also forms the basis for the decomposition of SRIRs that will be introduced in Sec.~\ref{sec:sub_dec_for_srirs}. In that context, the source signals $\bm s(t)$ represent the components of the direct part of the SRIR, i.e., direct sound and salient reflections, $\bm H$ describes their transfer paths to the microphone array, and $\bm n(t)$ represents the residual SRIR.}

The spatial covariance matrix $\bm R_\x$ is defined as the expectation $\E{\cdot}$ of the outer vector product,
\begin{equation}\label{eq:covariance-theory}
    \bm R_\x = \E{\bm x(t) \bm x(t)\T} \, .
\end{equation}
When assuming that the convolutively mixed signals $\bm H \bm s(t)$ and the noise $\bm n(t)$ are mutually uncorrelated, the covariance $\bm R_\x$ of the noisy signal is the sum of the covariance of the mixed source signals ${\bm R_\s = \E{\bm H \bm s(t) \bm s(t)\T \bm H\T}}$ and the covariance of the noise ${\bm R_\n = \E{\bm n(t) \bm n(t)\T}}$,
\begin{equation}\label{eq:covariance}
    \bm R_\x = \bm R_\s + \bm R_\n \, .
\end{equation}

If the noise is spatially white, i.e., it is uncorrelated across microphones and has common variance $\sigma_\n^2$, its covariance is a scaled identity matrix, ${\bm R_\n = \sigma_n^2 \bm I}$. The eigenvalue decomposition (EVD) of the array signal covariance then shows that the signal and noise covariance matrices share the same set of eigenvectors~\cite{Hermus2007}, which are collected in the columns of $\bm U$,
\begin{equation}\label{eq:covariance-evd}
    \bm R_\x = \bm U \bm \Lambda \bm U\T = \bm U (\bm \Lambda_\s + \sigma_\n^2 \bm I) \bm U\T\, .
\end{equation}
\rev{Thus, the eigenvalues of the noisy-signal covariance $\bm R_\x$ in the diagonal matrix $\bm \Lambda$ are equal to the eigenvalues of the source signal covariance $\bm R_\s$ in the diagonal matrix $\bm \Lambda_\s$ plus the noise variance $\sigma_\n^2$.} Note that due to the convolutive mixture and possible correlation between the source signals $\bm s(t)$, the rank $Q_\s$ of $\bm R_\s$ cannot be assumed to be equal to the number of sources $Q$. However, if the covariance $\bm R_\s$ is singular, i.e., its rank ${Q_\s < M}$, then $\bm \Lambda_\s$ contains zero-valued eigenvalues and the smallest ${Q_\n = M-Q_\s}$ eigenvalues of $\bm R_\x$ are equal to the noise variance $\sigma_\n^2$. 

This is the core observation that subspace methods exploit to reduce noise as it allows for a separation between eigenvalue-eigenvector pairs that are attributed to signal-plus-noise components and other pairs that are solely attributed to the noise. 
The eigenvectors corresponding to large eigenvalues are hence referred to as signal eigenvectors and are collected in the columns of $\bm U_\s$, while the eigenvectors corresponding to small eigenvalues equal to $\sigma_\n^2$ are referred to as noise eigenvectors and are collected in the columns of $\bm U_\n$. Noise reduction is then performed by an orthogonal projection of the noisy signal onto the signal subspace~\cite{Tufts1982},
\begin{equation}\label{eq:signal-subspace-signal}
    \bm x_\s(t) = \bm U_\s \bm U_\s\T \bm x(t) \, ,
\end{equation}
and the corresponding noise-only signal is obtained by an orthogonal projection onto the noise subspace,
\begin{equation}\label{eq:noise-subspace-signal}
    \bm x_\n(t) = \bm U_\n \bm U_\n\T \bm x(t) \, .
\end{equation}

\subsection{Subspace Decomposition Including a Noise Estimate}\label{sec:subspace-theory-with-noise-estimate}
If the noise $\bm n(t)$ is not spatially white, the noise covariance matrix $\bm R_\n$ is not a scaled identity matrix and the attribution of eigenvalues of the covariance $\bm R_\x$ to the signal or noise subspace based on their magnitudes fails. Note that, \rev{due to the finite distance between the diaphragms}, microphone array signals in rooms are never fully uncorrelated, not even in homogeneous diffuse fields~\cite[Ch.~2.2]{Herbordt2005}. However, if an estimate of the noise covariance matrix is available, the generalized eigenvalue decomposition (GEVD) of $\bm R_\x$ and $\bm R_\n$ diagonalizes the noisy-signal covariance $\bm R_\x$ and the noise covariance $\bm R_\n$ simultaneously~\cite{Roy1989a},
\begin{align}
    \bm \Psi\T \bm R_\x \bm \Psi &= \bm \Delta \, ,\label{eq:GEVD-x}\\
    \bm \Psi\T \bm R_\n \bm \Psi &= \bm I \, .\label{eq:GEVD-n}
\end{align}
The columns of $\bm \Psi$ contain the generalized eigenvectors and the diagonal matrix $\bm \Delta$ contains the generalized eigenvalues. As under the assumption of spatially white noise that was similarly exploited in~\eqref{eq:covariance-evd}, the generalized eigenvalues $\bm \Delta$ are equal to the eigenvalues of $\bm R_\s$ offset by one, $\bm \Delta = \bm R_\s + \bm I$. Thus, the generalized eigenvalues can be interpreted as the eigenvalues of a pre-whitened signal and the generalized eigenvectors span the signal subspace~\cite[Ch.~8.7]{Loizou2013}.
Note that as in the EVD, the eigenvectors are uniquely determined only up to an arbitrary factor. This factor is commonly chosen such that ${\bm \Psi\T \bm R_\n \bm \Psi = \bm I}$.

Due to the diagonalization of the noise covariance, a magnitude-based discrimination between signal and noise eigenvalues is possible again but the generalized eigenvectors in the columns of $\bm \Psi$, which are the eigenvectors of ${\bm R_\n^{-1} \bm R_\x}$, are not orthogonal as ${\bm R_\n^{-1} \bm R_\x}$ is not necessarily symmetric. The projection onto signal and noise subspace is hence expressed as~\cite{Hu2003a},
\begin{align}
    \bm x_\s(t) &= \bm \Psi^{-\mathrm{T}} \bm \Gamma_\s \bm \Psi\T \bm x(t) \, ,  \label{eq:GEVD-xs}\\
    \bm x_\n(t) &= \bm \Psi^{-\mathrm{T}} \bm \Gamma_\n \bm \Psi\T \bm x(t) \, , \label{eq:GEVD-xn}
\end{align}
where the diagonal binary selection matrices ${\bm \Gamma_\s = \bm I_{Q_\s\times M}\T \bm I_{Q_\s\times M}}$ and ${\bm \Gamma_\n = \bm I - \bm \Gamma_\s}$ contain $Q_\s$ ones as the first $Q_\s$ diagonal entries and ${Q_\n = M-Q_\s}$ ones as the last $Q_\n$ diagonal entries, respectively, and zeros otherwise, to select the $Q_\s$ and $Q_\n$ eigenvectors of their corresponding subspace. As the sum of $\bm \Gamma_\s$ and $\bm \Gamma_\n$ is the identity matrix, the sum of the signal and noise subspace signals perfectly reconstructs the original array signal, $\bm x(t) = \bm x_\s(t) + \bm x_\n(t)$. The subspace decomposition using the GEVD is mathematically equivalent to the sequence of pre-whitening the signal, decomposing it into the subspaces, and de-whitening the result~\cite[Ch.~8.7]{Loizou2013}. 

If the noise estimate is obtained by discrete measurements of the noise-only signal, the explicit computation of the signal and noise covariances can be avoided for computational efficiency by employing the generalized singular value decomposition (GSVD) instead of the GEVD~\cite{Doclo2002b}. The GSVD relies on data matrices, i.e., matrices containing multiple observations of the signal. Let $\bm X$ and $\bm N$ be ${K\times M}$ and ${L \times M}$ matrices that contain $K$ and $L$ observations of their corresponding signals $\bm x(t)$ and $\bm n(t)$. 
The GSVD then decomposes the noisy-signal data matrix $\bm X$ and the noise data matrix $\bm N$ into an orthogonal matrix $\bm V_\x$ or $\bm V_\n$, a non-negative diagonal matrix $\bm \Sigma_\x$ or $\bm \Sigma_\n$, and a common square matrix $\bm \Phi$,
\begin{align}
	\bm X &= \bm V_\x \bm \Sigma_\x \bm \Phi\T \, , \label{eq:gsvd-of-signal-matrix}\\
	\bm N &= \bm V_\n \bm \Sigma_\n \bm \Phi\T \, . \label{eq:gsvd-of-noise-matrix}
\end{align}
The matrices $\bm \Sigma_\x$ and $\bm \Sigma_\n$ contain the singular values, while $\bm V_\x$ and $\bm V_\n$ contain the respective left singular vectors and $\bm \Phi$ contains the common right singular vectors. For notational brevity, we assume an \emph{economy-sized} GSVD and ${K\geq M}$, ${L \geq M}$, so that $\bm \Sigma_\x$ and $\bm \Sigma_\n$ are ${M \times M}$ square matrices.

The generalized eigenvalues of $\bm R_\x$ and $\bm R_\n$ on the diagonal of $\bm \Delta$ are obtained via the GEVD, cf.~\eqref{eq:GEVD-x}. If the corresponding covariance matrices are estimated via the sample covariance, i.e., ${\bm R_\x = \frac{1}{K} \, \bm X\T \bm X}$ and ${\bm R_\n = \frac{1}{L} \, \bm N\T \bm N}$, the generalized eigenvalues $\bm \Delta$ can equivalently be obtained via the GSVD~\cite[Ch.~8.7.4]{Golub2013},
\begin{equation}
	\bm \Delta = \frac{L}{K} \, (\bm \Sigma_\x\T \bm \Sigma_\x) (\bm \Sigma_\n\T \bm \Sigma_\n)^{-1} \, .
\end{equation}
For convenience, we further define the vector
\begin{equation}
    \bm \sigma = \dg{(\bm \Sigma_\x\T \bm \Sigma_\x) (\bm \Sigma_\n\T \bm \Sigma_\n)^{-1}} \, ,
\end{equation}
that contains the squared generalized singular values (GSVs) in decreasing order. The $\dg{\cdot}$ operator transfers the entries from the main diagonal of a matrix to a vector. The squared GSVs are an essential part of the proposed threshold selection mechanism in Sec.~\ref{sec:thresholding} and will simply be referred to as GSVs in the following.

In the case of the GSVD, the subspace decomposition is performed as a low-rank approximation, resulting in the ${K \times M}$ signal subspace and noise subspace matrices $\bm X_\s$ and $\bm X_\n$~\cite[Ch.~8.4]{Loizou2013}. Similar to the subspace decomposition using the GEVD in~\eqref{eq:GEVD-xs} and~\eqref{eq:GEVD-xn}, the $Q_\s$ largest singular values and their corresponding singular vectors are used to obtain the signal subspace components and the last ${Q_\n = M-Q_\s}$ singular vectors corresponding to the smallest singular values are used to obtain the noise subspace components,
\begin{align}
    \bm X_\s &= \bm V_\x \bm \Sigma_\x \bm \Gamma_\s \bm \Phi\T \, , \\
    \bm X_\n &= \bm V_\x \bm \Sigma_\x \bm \Gamma_\n \bm \Phi\T \, .
\end{align}

Note that some subspace methods in the literature impose more assumptions on the signals and thus are able to gain more information in the decomposition process. They typically either assume narrow-band signals~\cite{Schmidt1986, Roy1989a} or assume short-term stationary signals and perform blockwise processing in the frequency domain~\cite[Ch.~5.2]{VanTrees2002}. These assumptions reduce the convolutive MIMO signal model in~\eqref{eq:MIMO_sig_model} to a multiplicative model, i.e., the FIR filters in $\bm H$ reduce to single-sample scaling factors. The rank $Q_\s$ of the source covariance $\bm R_\s$ is then equal to the number of sources $Q$ and the individual source signals are decorrelated which facilitates a source parameter estimation. However, as established in Sec.~\ref{sec:sub_dec_for_srirs}, the assumptions that are imposed on the signals in this work are more restrictive.

\section{Subspace Decomposition of Spatial Room Impulse Responses}\label{sec:sub_dec_for_srirs}
\subsection{Signal Model}
Motivated by their perceptual relevance, the proposed algorithm aims at separating the direct sound and salient reflections from the SRIR. Applying the nomenclature from Sec.~\ref{sec:subspace_theory} to the present context means that direct sound and salient reflections are considered convolutively mixed source signals and everything else is considered noise. In the following, the direct sound and salient reflections will also be referred to as the direct part or direct subspace components, and the rest, containing the superposition of an increasing amount of reflections and noise will be referred to as the residual.

\rev{The spatio-temporal properties of the residual of the SRIR typically change over time}. In the early part of the SRIR, the residual mainly contains noise and non-transient components of the room response due to room modes. As time progresses, it is additionally comprised of a superposition of non-salient reflections. Toward the later part of the SRIR, the residual is dominated by the superposition of exponentially increasing numbers of non-salient reflections. This reverberation might exhibit isotropic or anisotropic properties, or a combination of both that varies over time~\cite{Gover2004, Alary2021}. During the late part of the SRIR, no salient reflections are expected so that the SRIR is composed of only the residual and no direct part.

To adapt to the \rev{spatio-temporal variations} of the residual, we propose to update the residual estimate whenever no salient reflections are detected within a signal block. The procedure is assumed to be successful if the properties of the residual \rev{change slower than the residual estimate is updated}. Additionally, we propose to process the SRIR backward in time to be able to obtain a reliable residual estimate before any salient reflection occurs. \rev{This process is illustrated in Fig.~\ref{fig:big-fig} and a more detailed overview of the algorithm will be given in Sec.~\ref{sec:algo_overview}}.

The covariance matrix is assumed to be estimated via the sample covariance of a signal block that contains a limited number of signal observations, cf.~Sec.~\ref{sec:subspace-theory-with-noise-estimate}. However, when the GSVD is used, the sample covariance is not calculated explicitly. As the direct sound and the salient reflections are highly transient signals, the rank of the source covariance matrix depends on the correlation of the captured signals and the temporal separation of the transients that are induced by a single or multiple reflections. The correlation depends on the transfer function from the source to the individual microphones, i.e., on properties of the acoustic environment and the array, and the temporal separation depends on the distance between the microphones.

The only requirement for the microphone array is that it has a small aperture so that sound pressures that are generated by a reflection are captured by a single signal block. \rev{Determining the duration of a reflection is not a straightforward task as it is influenced by the array aperture, diffraction, and scattering. However, the propagation delay of a sound wave across the maximum array dimension is often a good approximation.} In practice, often spherical arrays are employed and their signals are transformed to the spherical harmonic (SH) domain, where they are radial filtered to compensate for the array radius and the scattering of the array baffle~\cite[Ch.~2.6]{Rafaely2019}. This leads to the typical assumption that spherical arrays in SH-domain processing have frequency-independent steering vectors so that, as with narrow-band signals, a multiplicative signal model is sufficient. However, the necessary regularization of the radial filters and spatial aliasing in practice limit this property to a narrow frequency region~\cite{Deppisch2021b}. Thus, these assumptions are not made in the proposed broadband algorithm so that it can either be directly applied to the microphone signals or to an SH decomposition thereof. We demonstrate the application of the proposed algorithm with both signal representations in \rev{Sec.~\ref{sec:space_vs_sh_domain}}.

\subsection{Rank Analysis of the Covariance Matrix}
\begin{figure*}
\centering
\subfloat[]{\includegraphics[width=\columnwidth]{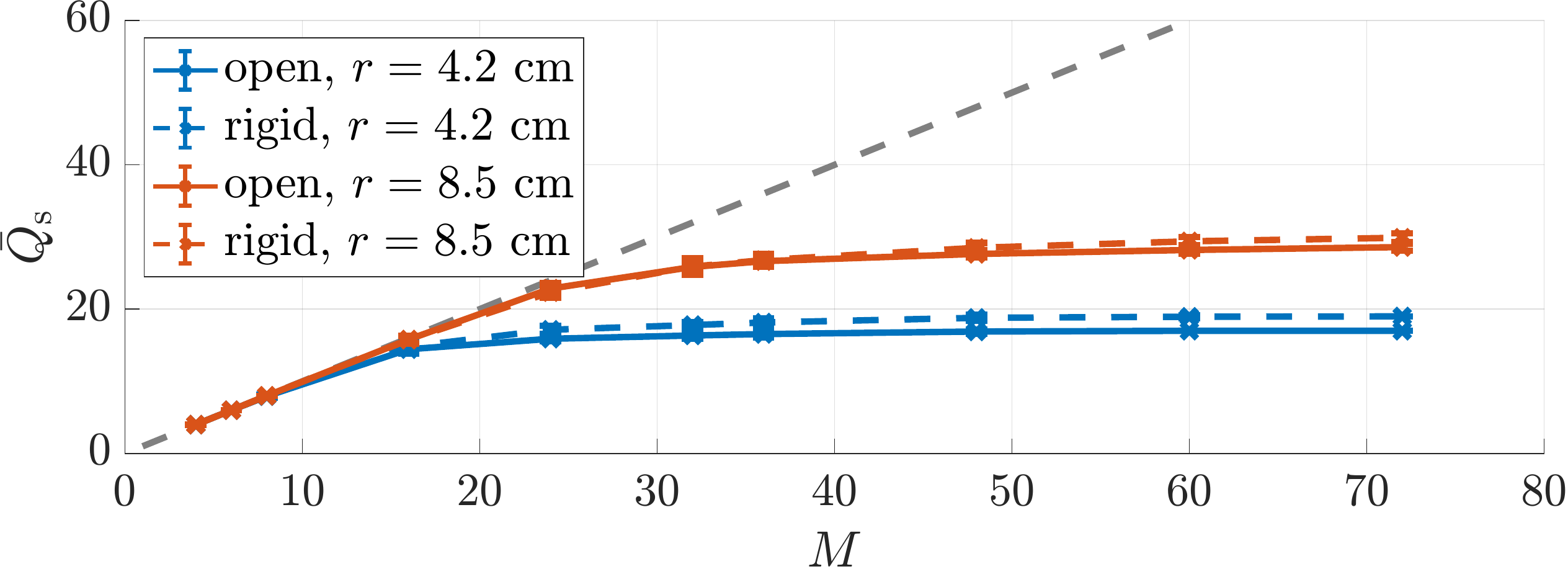}}
\hfill
\subfloat[]{\includegraphics[width=\columnwidth]{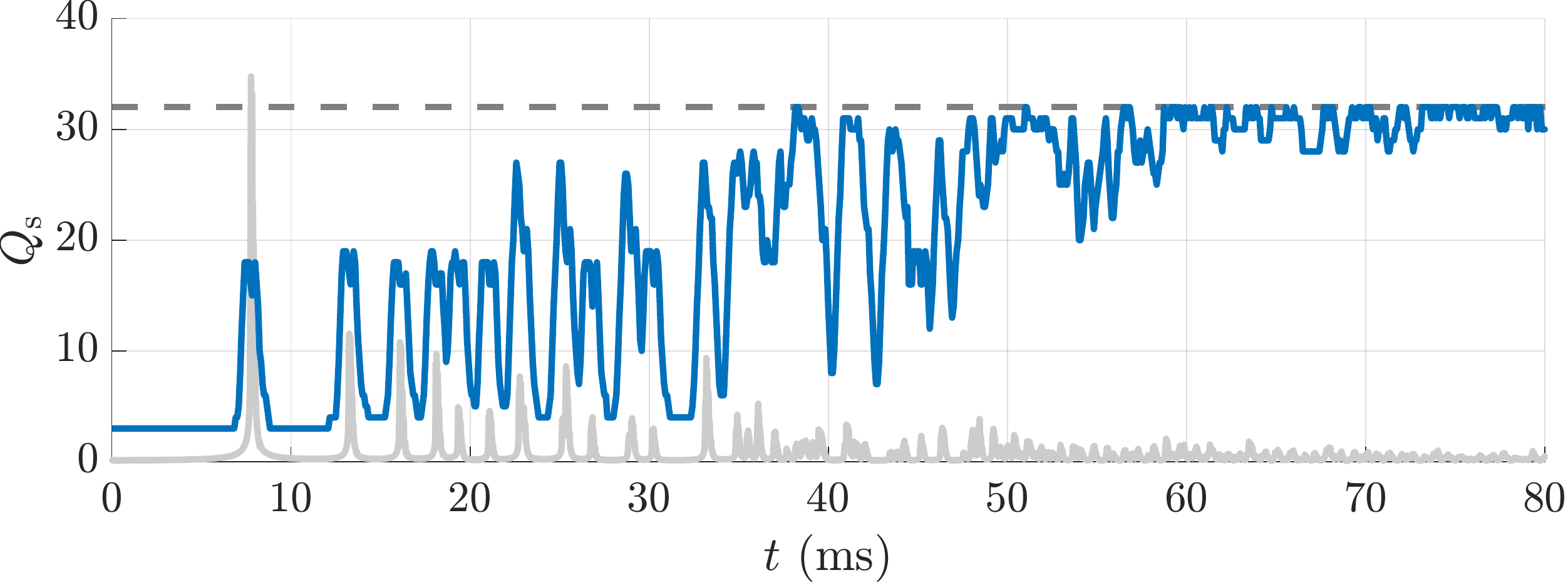}}
\vspace{-0.2cm}
\caption{The subspace decomposition can be performed if the source covariance matrix $\bm R_\s$ is singular, i.e., its rank ${Q_\s < M}$. \textbf{(a)} \rev{The mean rank $\bar{Q}_\s$ due to a single impinging plane wave depends on the number of microphones $M$, on the array radius $r$, and on the array surface being rigid or open. It is singular if it stays below the dashed gray line illustrating the number of microphones. \textbf{(b)} The source covariance matrix of an SRIR was simulated using the image source method. It is singular in the early part}. The summed magnitude of the SRIR is shown in gray for reference.}
\label{fig:rankAnalysis}
\end{figure*}
The core assumption that facilitates noise reduction via subspace methods is that the source signals, which are in the present case the direct sound and the salient reflections, only occupy a subspace of the full signal space. In other words, noise reduction is only possible if the source covariance matrix $\bm R_\s$ is singular, ${Q_\s < M}$. To determine if a subspace decomposition is feasible for salient reflections in an SRIR, in the following, the rank of the source covariance matrix is analyzed, first for an individual plane wave impinging on different microphone arrays and then for a simulated SRIR.

Fig.~\ref{fig:rankAnalysis}~(a) shows the mean and the standard deviation of the rank $Q_\s$ of the covariance matrix for a plane wave that impinges on different spherical microphone arrays under anechoic conditions. Note that the standard deviations are small and thus hardly visible. The dashed gray line illustrates the maximum rank $M$, which is equal to the number of microphones in the array. The mean rank $\bar{Q}_\s$ is obtained by averaging the rank over 240 incidence directions that are distributed according to a t-design~\cite{Hardin1996a}. The rank for each incidence direction is calculated as the number of eigenvalues of the covariance matrix that are less than \SI{100}{dB} below the largest eigenvalue. The plane waves and the spherical scattering were simulated using the spherical microphone array impulse response generator (SMIRGen)~\cite{Jarrett2012}. The simulated array configurations include 10 different spherical arrangements comprising between 4 and 72~microphones that all are arranged according to t-designs. All 10~arrangements were simulated as open and rigid arrays, and with radii of \SI{4.2}{cm} and \SI{8.5}{cm}.

The covariance matrix on average has full rank for the array configurations that comprise 4, 6, and 8~microphones. The mean rank $\bar{Q}_\s$ is close to being full in the case of the array configurations with 16 microphones and in the case of the arrays with the larger radius and 24 microphones. For all other configurations, the mean rank is clearly singular. Thus, a subspace decomposition for a single impinging plane wave can be performed with the simulated arrays of radius \SI{4.2}{cm} that comprise ${M=24}$ or more microphones and with the arrays of radius \SI{8.5}{cm} that comprise ${M=32}$ or more microphones. With an increasing number of microphones $M$, all tested configurations converge to a maximum rank that depends on the array configuration and is equal to 29 and 30 for the open and rigid sphere configurations with a radius of \SI{8.5}{cm}, and equal to 17 and 19 for the configurations with a radius of \SI{4.2}{cm}. 
The covariance matrices from microphone arrays with the larger radius generally have a higher rank $Q_\s$ than the ones from the smaller arrays since their signals contain larger temporal delays and since they capture less-correlated signals because those arrays are large compared to the wavelength down to lower frequencies. \rev{The observed rank is also slightly higher in the case of rigid arrays because the scattering of sound waves off their surface additionally decorrelates the captured signals.}

To determine if a subspace decomposition is still feasible if more than one reflection occurs per analysis window, we analyze the evolution of the rank $Q_\s$ for an SRIR. The SRIR was generated using the image source method~\cite{Allen1979} and SMIRGen, assuming a shoe-box room of dimensions $8 \times 7 \times 6$~\si{m}. \rev{Note that the goal of this simulation is not to render a highly-realistic SRIR but to investigate the rank of the direct-part (source) covariance matrix $\bm R_\s$ due to the direct sound and individual reflections, not the full covariance $\bm R_\x$ that will be used in the subspace decomposition. The separation of eigenvalues can be attempted as in~\eqref{eq:covariance-evd} only if the source covariance $\bm R_\s$ is singular.} The simulated array is of radius \SI{4.2}{cm} and comprises 32~microphones that are arranged according to a t-design. Fig.~\ref{fig:rankAnalysis}~(b) shows $Q_\s$ during the first \SI{80}{ms} of the SRIR. The summed magnitude of the SRIR is shown in gray for reference. The sample covariance matrix was calculated in 32-sample (\SI{0.7}{ms}) rectangular windows with a hop size of 4~samples. The rank $Q_\s$ was calculated as the number of eigenvalues of the covariance matrix that are less than \SI{100}{dB} below the largest eigenvalue in each window.

During the first \SI{38}{ms}, the rank $Q_\s$ consistently stays below the maximum possible rank of ${M=32}$, which is again shown as a dashed gray line. Between \SI{38}{ms} and \SI{59}{ms}, $Q_\s$ fluctuates over a wide range of values and approaches the maximum rank multiple times. After \SI{59}{ms}, the rank stays close to the maximum rank. Hence, for the given SRIR a subspace decomposition can separate the direct part from the residual in the early part of the SRIR until \SI{38}{ms} and might be able to separate some salient reflections until \SI{59}{ms}. These specific time spans do not generalize to other SRIRs and microphone arrays. However, it can be assumed that the time span where a separation is effective increases with an increasing number of microphones and that the subspace decomposition is effective in the early part of typical SRIRs if a suitable microphone array is used. This is also shown in the case study in Sec.~\ref{sec:case_study_measured_srirs}, where the proposed method is applied to three measured SRIRs with different acoustic properties. \rev{In practice, the decomposition of the SRIR might be mainly relevant up to the perceptual mixing time~\cite{Lindau2010} and thus the method could be limited to the early part of the SRIR according to a mixing time estimate.}

\subsection{Algorithm Overview}\label{sec:algo_overview}
\begin{figure*}[t]
\centering
\subfloat[]{\begin{tikzpicture}[>=stealth]
    \node[anchor=south west,inner sep=0] (image) at (0,0) {\includegraphics[width=0.8\textwidth, trim=1cm 0cm 1cm 2cm, clip]{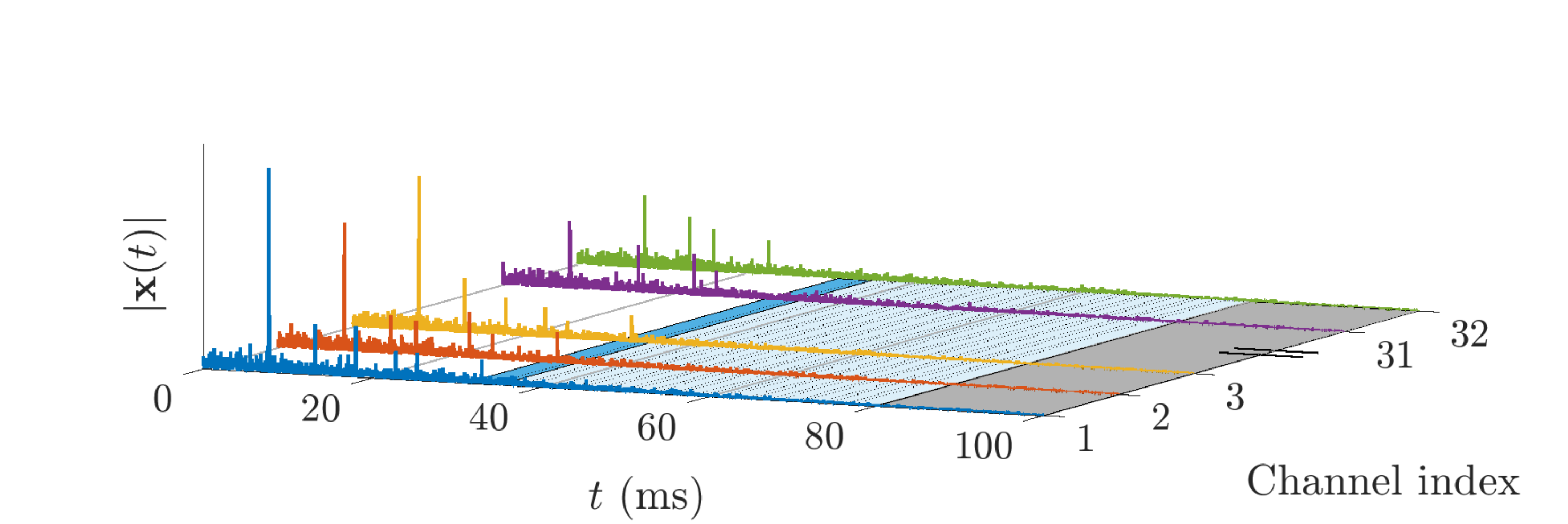}};
    \begin{scope}[x={(image.south east)},y={(image.north west)}]
        \draw[thick] (0.83,0.48) -- (0.87,0.85) node[above] {\footnotesize 1.) Initial residual estimate};
        \draw[thick, ->] (0.8,0.58) -- (0.6, 0.63) node[above right, text width=4cm] {\footnotesize 2.) Blockwise GSVD,\\update residual estimate};
        \draw[thick] (0.53, 0.55) -- (0.45, 0.85) node[above] {\footnotesize 3.) Threshold exceeded: subspace decomposition};
    \end{scope}
\end{tikzpicture}}
\vspace{-0.4cm}
\subfloat[]{\includegraphics[width=0.5\textwidth, trim=1cm 0cm 0cm 2cm, clip]{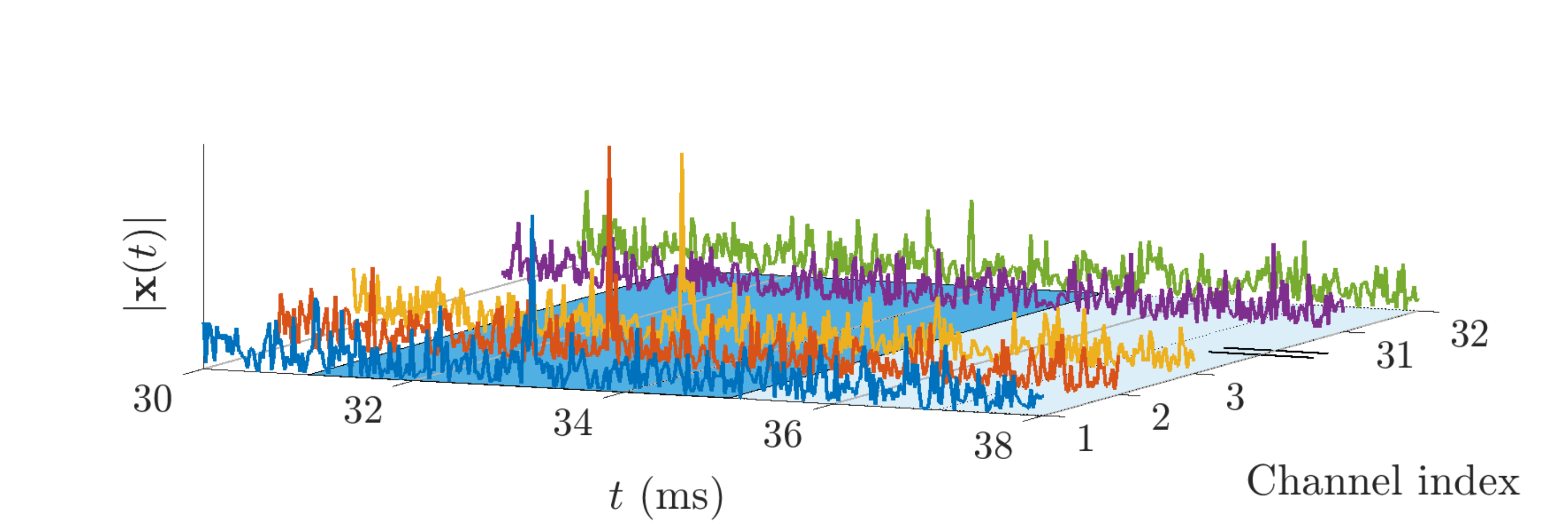}%
\label{fig:rir-zoomed}}
\hfill
\subfloat[]{\includegraphics[width=0.5\textwidth, trim=1cm 0cm 0cm 2cm, clip]{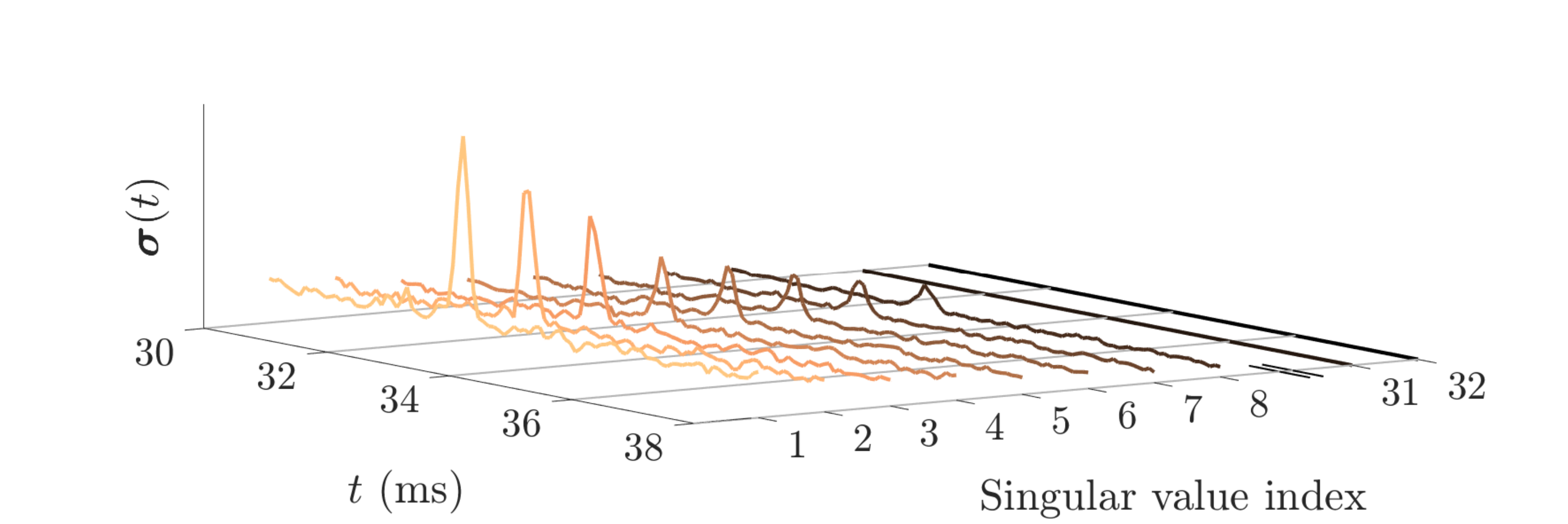}%
\label{fig:evals-zoomed}}\\
\vspace{-0.4cm}
\subfloat[]{\includegraphics[width=0.5\textwidth, trim=1cm 0cm 0cm 2cm, clip]{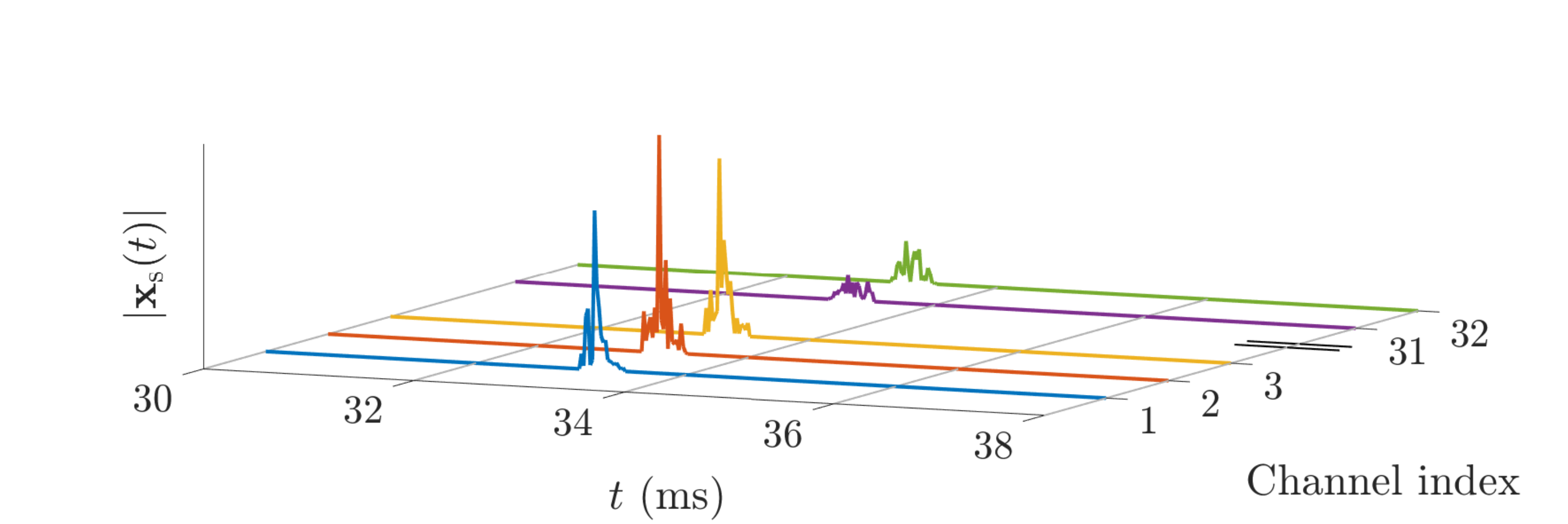}%
\label{fig:dir-zoomed}}
\hfill
\subfloat[]{\includegraphics[width=0.5\textwidth, trim=1cm 0cm 0cm 2cm, clip]{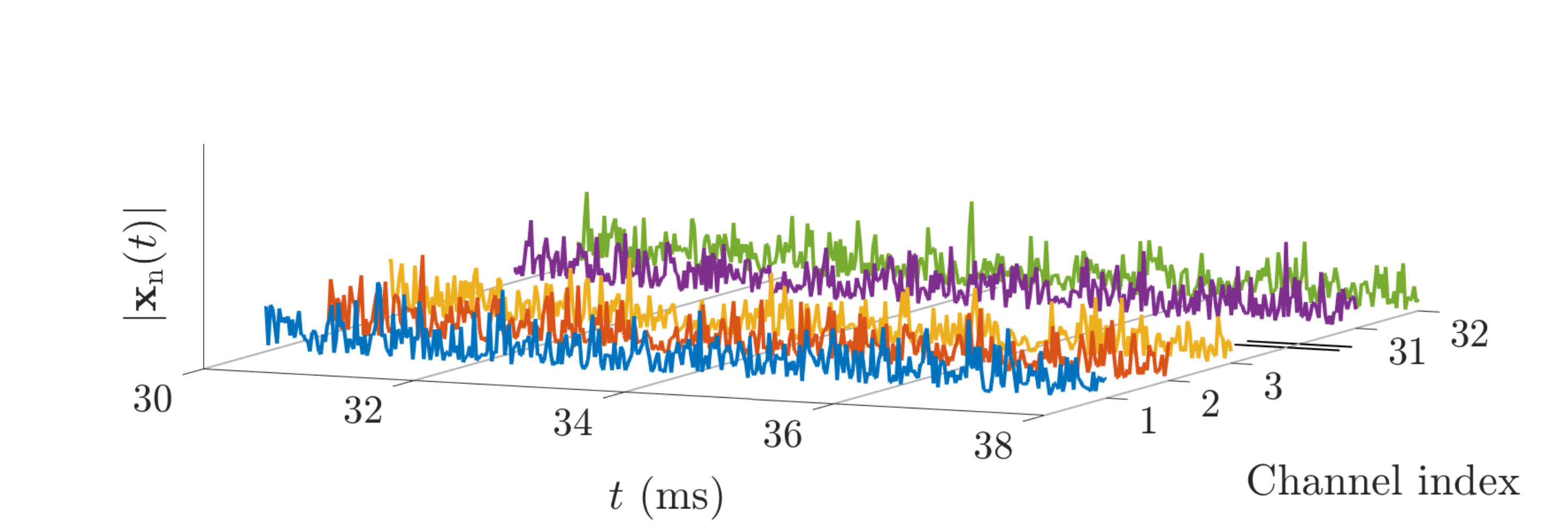}%
\label{fig:diff-zoomed}}
\vspace{-0.1cm}
\caption{Direct and residual subspace decomposition of a 32-channel SRIR $\bm x(t)$. \textbf{(a)} The proposed algorithm first takes an initial residual estimate from the end of the SRIR. It then proceeds toward the beginning of the SRIR and performs the GSVD on every signal block. If the sum of the GSVs $\bm \sigma(t)$ is below the detection threshold, the residual estimate is updated. If their sum exceeds the threshold, the subspace decomposition is performed. \textbf{(b)} A zoomed-in part of the SRIR contains a salient reflection. \textbf{(c)} The eight largest GSVs of the zoomed-in part exhibit a distinct peak at the location of the reflection. The two smallest GSVs do not exhibit a visible peak. \textbf{(d)} The direct signal $\bm x_\s(t)$ contains the salient reflection from (b). \textbf{(e)} The residual signal $\bm x_\n(t)$ does not contain the reflection.}
\label{fig:big-fig}
\end{figure*}
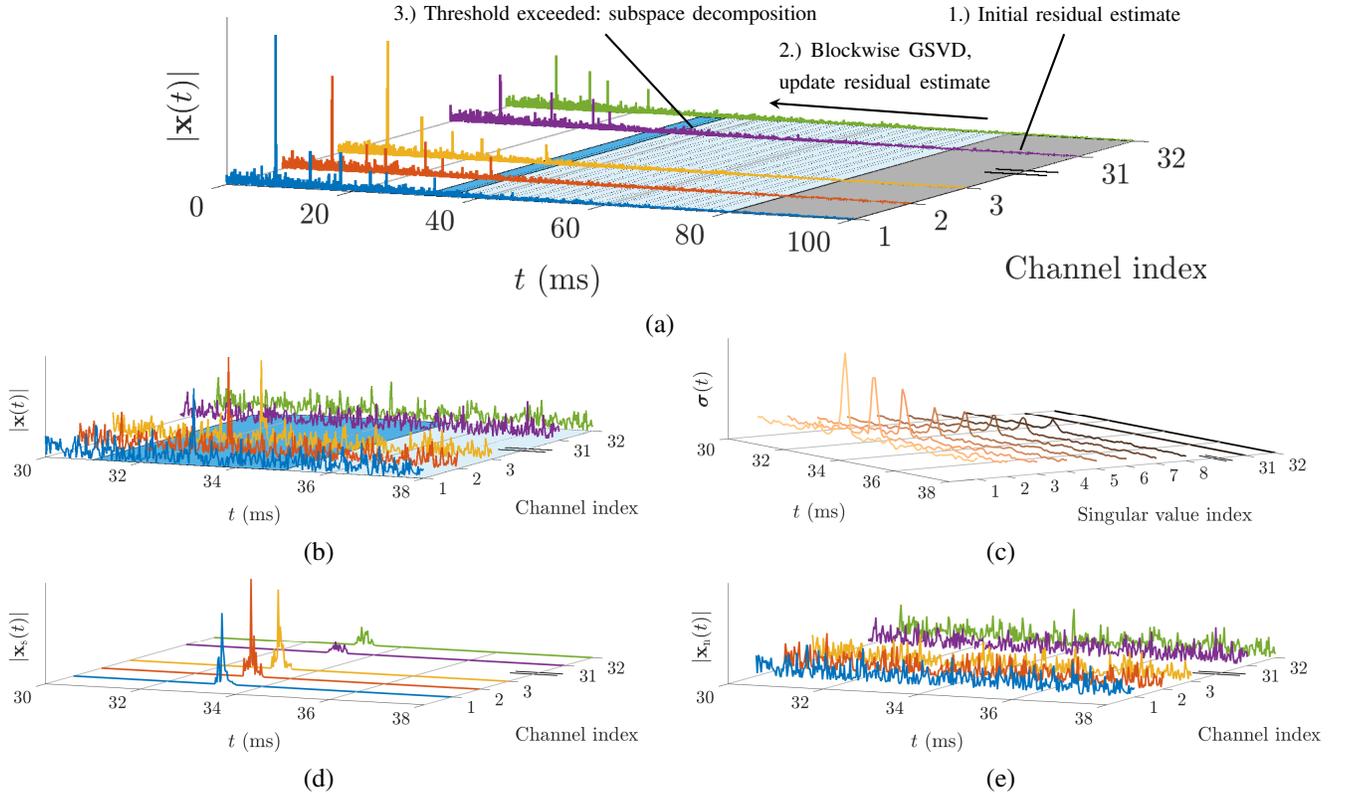

Fig.~\ref{fig:big-fig} illustrates the application of the proposed subspace decomposition algorithm to a simulated SRIR. The SRIR is simulated as direct sound and first-order image-source reflections in exponentially decaying noise. For convenience and without loss of generality, the decaying noise starts before the direct sound occurs. The simulated, rigid spherical microphone array is the same that was used in Fig.~\ref{fig:rankAnalysis}~(b). It has a radius of \SI{4.2}{cm} and comprises 32 microphones that are arranged according to a spherical \rev{t-design of degree~7~\cite{Hardin1996a}}. Again, the image-source method was employed using SMIRGen~\cite{Jarrett2012}. The microphone signals of the exponentially decaying noise tail were generated to exhibit the spatial coherence of the array in an isotropic spherical noise field using the method from~\cite{Habets2008}.

Fig.~\ref{fig:big-fig}~(a) shows five channels of the full 32-channel SRIR. The algorithm first takes a signal block from the end of the SRIR as an initial residual estimate and then performs a blockwise GSVD while proceeding toward the beginning of the SRIR. If the sum of the generalized singular values (GSVs) exceeds the detection threshold, the SRIR is decomposed into the direct part and the residual. Otherwise, only the residual estimate is updated. 

Fig.~\ref{fig:big-fig}~(b) shows a magnified section of the SRIR that includes a salient reflection. A subset of the corresponding GSVs is shown in Fig.~\ref{fig:big-fig}~(c). The first eight GSVs exhibit a distinct peak at the location of the reflection while the smallest two GSVs do not exhibit a visible peak. All $Q_\s$ GSVs above a given threshold are attributed to the reflection whereas the $Q_\n=32-Q_\s$ smaller GSVs are attributed to the residual. In the present case, $Q_\s$ was chosen to be 6. A method that determines this threshold is proposed in Sec.~\ref{sec:thresholding}. 
The direct part SRIR that contains the reflection is shown in Fig.~\ref{fig:big-fig}~(d) and the residual SRIR in Fig.~\ref{fig:big-fig}~(e). The sum of the two reconstructs the original SRIR.

\subsection{Threshold Selection}\label{sec:thresholding}
The selection of appropriate thresholds for the detection of salient reflections and the estimation of the number of direct subspace components $Q_\s$ is key to a successful decomposition. Common criteria to find the number of signal subspace components are either based on the ratio of the geometric mean to the arithmetic mean of a number of small eigenvalues~\cite{Wax1985}, in a nutshell rating the \emph{equality} of a subset of eigenvalues, or using measures to find the gap between a set of larger and a set of smaller eigenvalues~\cite{He2010,Cong2012}. The methods are based on the assumption of eigenvalues of the noise subspace being similar in size and do not exploit prior information like a noise estimate.

To facilitate a robust threshold selection by incorporating information from the residual estimate, we propose a threshold measure based on the cumulative sum of the GSVs. It is inspired by~\cite{Bakamidis1991}, where the number of components $Q_\s$ is selected so that the reconstruction error is close to an estimate of the noise variance. 
The Frobenius norm of a matrix is the root of the sum of its squared elements and is equal to the root of the sum of its squared singular values~\cite[Ch.~2.4.2]{Golub2013}. In the case of data matrices containing microphone array signals, the Frobenius norm can be interpreted as the energetic sum of all microphone signals. A threshold that keeps the total energy of the residual in the presence of salient reflections equal to the energy of the full signal in the absence of salient reflections can hence be defined via the sum of squared singular values. 

The proposed threshold is based on this idea, however, two further observations lead to its precise definition: i)~In contrast to the orthogonal left and right singular vectors of the SVD, the GSVD involves the non-orthogonal right singular vectors $\bm \Phi$, cf.~\eqref{eq:gsvd-of-signal-matrix}. In consequence, the rooted sum of the squared singular values in $\bm \Sigma_\x$ is not equal to the Frobenius norm of the signal matrix $\bm X$. However, the GSVs $\bm \sigma$ can be interpreted as the singular values of the pre-whitened signal and the SVD of the pre-whitened signal, which is not explicitly calculated, involves orthogonal singular vectors. Thus, by choosing the number of residual components $Q_\n$ during the decomposition so that the sum of the corresponding $Q_\n$ GSVs equals the sum of all GSVs in the absence of salient reflections, the energy of the whitened residual can be kept constant. 

ii)~While the residual subspace only contains residual components, the direct subspace contains a superposition of direct and residual components. The singular values of the whitened signal are the GSVs and they all carry equal parts of the variance of the whitened residual. Thus, the sum of the $Q_\n$ GSVs during the decomposition needs to be smaller than the sum of all GSVs in the absence of salient reflections. More precisely, if $Q_\n$ GSVs are attributed to the residual, they should carry a fraction of $Q_\n/M$ of the full energy that is determined when no salient reflections are present. Recall that $M$ denotes the number of microphones.

\begin{figure*}
\centering
\subfloat[]{\includegraphics[width=\columnwidth]{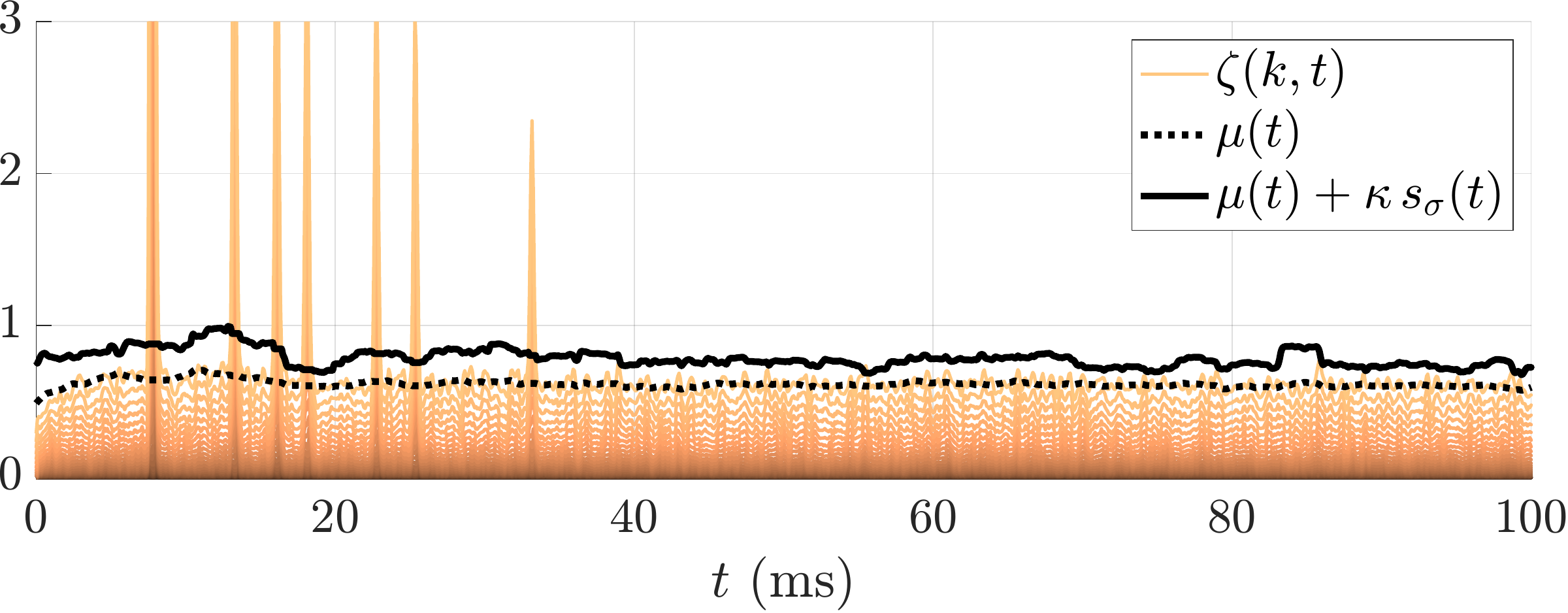}}
\hfill
\subfloat[]{\includegraphics[width=\columnwidth]{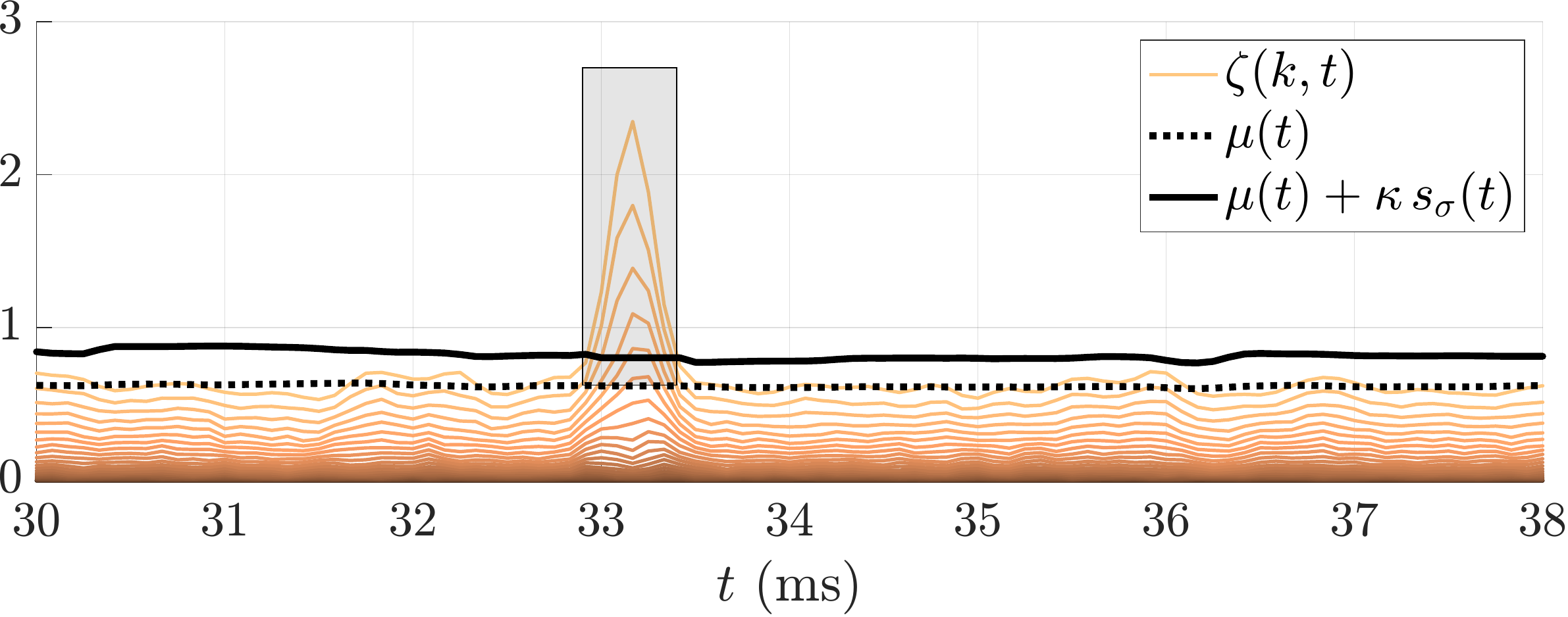}}
\vspace{-0.2cm}
\caption{The weighted cumulative sums $\zeta(k,t)$ of the GSVs, shown in shades from orange to brown, exhibit distinct peaks at times where reflections occur. \textbf{(a)} The GSV sum, which is the largest of the cumulative sums $\zeta(k,t)$, exceeds the detection threshold, drawn as a solid black line, for the direct sound and each of the 6 reflections. Thus, all direct components are detected. \textbf{(b)} Zoomed-in section of (a) around a reflection. The left and right borders of the gray rectangle mark the time instances where a reflection is detected. The number of direct subspace components is determined as the number of weighted, cumulatively summed GSVs $\zeta(k,t)$ that exceed the time-averaged sum $\mu(t)$ of the GSVs, which is shown as a dotted black line. At the peak, this results in 6 direct components and 26 residual components.}
\label{fig:thresholding}
\end{figure*}
In mathematical terms, the proposed criterion determines the number of residual components $Q_\n$ as the maximum integer $k$ for which the cumulative sum of $k$ GSVs $\zeta(k,t)$ is still smaller than the time-averaged sum of the GSVs $\mu(t)$ in the absence of salient reflections,
\begin{equation}\label{eq:Q_threshold_averaged}
    Q_\n(t) = \max(\lbrace k \in \mathbb{N}^+ : \zeta(k,t) < \mu(t) \rbrace) \, ,
\end{equation}
where
\begin{equation}
    \zeta(k,t) = \frac{M}{k}\, \sum_{m=M-k+1}^{M} [\bm \sigma(t)]_m \, 
\end{equation}
is the cumulative sum of the $k$ smallest GSVs weighted by ${M/k}$ and $[\bm \sigma(t)]_m$ denotes the $m$-th element of the vector $\bm \sigma(t)$, i.e., the $m$-th largest GSV. The weighting $M/k$ stems from the above reasoning that showed that the whitened-residual energy should be a ${Q_\n / M}$ part of the GSV sum. As $Q_\n$ is being determined at this point, it has been replaced by the index $k$. The fraction $k/M$ has further been transferred to the left side of the inequality in~\eqref{eq:Q_threshold_averaged} to become ${M/k}$. The time-averaged GSV sum
\begin{equation}
    \mu(t) = \mathcal{F}_{\text{CMA}} \left\{ \sum_{m=1}^M [\bm \sigma(t)]_m \right\} \, 
\end{equation}
is obtained by applying the constrained moving average filter $\mathcal{F}_{\text{CMA}} \{\cdot \}$ that time-averages the sum of the GSVs and is only updated if no salient reflection is present.
Once the number of residual components $Q_\n$ is known, the number of direct subspace components is obtained as ${Q_\s = M - Q_\n}$.

A second thresholding mechanism is needed to detect the presence of reflections. Only if reflections are detected, the estimation of the number of subspace components is performed. Similar to the estimator for the number of subspace components, the proposed detection threshold is based on the sum of the GSVs that is averaged over time instants without salient reflections. Reflections are detected if the sum of the GSVs of the current observation is larger than the time-averaged sum of previous GSVs $\mu(t)$ plus a multiple $\kappa$ of their standard deviation $s_{\sigma}(t)$,
\begin{equation}
    \sum_{m=1}^M [\bm \sigma(t)]_m > \mu(t) + \kappa \,  s_{\sigma}(t) \, .
\end{equation}
If that concept is to be implemented, a first-in-first-out (FIFO) buffer that contains a number of observations of the sum of the GSVs and is only updated during time instances without salient reflections is beneficial. The averaged GSV sum $\mu(t)$ and the standard deviation $s_{\sigma}(t)$ are then calculated as the arithmetic mean and the standard deviation of all observations in the buffer.

Fig.~\ref{fig:thresholding} shows the weighted cumulative GSV sums $\zeta(k,t)$ in shades from orange to brown, the time-averaged sum of GSVs $\mu(t)$ as a dotted black line, and the reflection detection threshold ${\mu(t) + \kappa \, s_{\sigma}(t)}$ as a solid black line for the same simulated SRIR as in Fig.~\ref{fig:big-fig}. The detection threshold is calculated using ${\kappa = 4}$. \rev{The parameter selection process is further described in Sec.~\ref{sec:param_influence} and in the supplementary material referenced therein.} Due to the iterative update of the residual estimate, the GSVs and also their cumulative sums stay constant in the absence of reflections although the reverberation is exponentially decaying, see Fig.~\ref{fig:big-fig}~(a). Reflections are detected whenever the sum of the GSVs, which is equivalent to the largest cumulative sum, is larger than the detection threshold. As shown in Fig.~\ref{fig:thresholding}~(a), the GSV sum exceeds the detection threshold for the direct sound and all 6 reflections and hence all direct components are detected.

Fig.~\ref{fig:thresholding}~(b) shows a zoomed-in section around the occurrence of the last reflection. The gray rectangle illustrates the estimation of the number of direct subspace components $Q_\s$. The left and right boundaries of the rectangle illustrate the temporal bounds in which the GSV sum is larger than the detection threshold. Within these bounds, the subspace decomposition is performed. The number of direct components is the number of weighted, cumulatively summed GSVs $\zeta(k,t)$ that is larger than the averaged GSV sum $\mu(t)$, resulting in ${Q_\s=6}$ direct components and ${Q_\n=26}$ residual components at the peak location.

\subsection{Influence of the Parameters}\label{sec:param_influence}
This section discusses the influence of the different parameters including the block size, the detection threshold offset $\kappa$, the amount of GSV averaging for the calculation of the thresholds, and the length of the residual estimate. In practice, the optimal parameter values depend on the acoustic environment and the employed microphone array, and they can be found by analyzing the decomposition results, the evolution of GSV sums, and the proposed thresholds as in Fig.~\ref{fig:thresholding}. For brevity, we discuss the influence of the parameters here and provide examples that illustrate the influence of the different parameters and the parameter selection process as supplementary material\footnote{A MATLAB Live Script and a corresponding PDF document are provided at \url{https://github.com/thomasdeppisch/SRIR-Subspace-Decomposition}.}.

The block size determines the temporal resolution of the subspace decomposition. A lower bound of the block size in samples is given by the number of microphones (or SH coefficients) of the employed array to be able to exploit the full signal space. Additionally, the block size should capture the full propagation delay of a sound wave across the maximum array dimension. For instance, in the case of the spherical Eigenmike em32 array with a radius of \SI{4.2}{cm}, we assume a maximum dimension of \SI{8.4}{cm} and a corresponding propagation delay of \SI{0.24}{ms}. Larger block sizes decrease the temporal resolution of the calculated GSVs and thus may reduce the temporal precision of the extraction of salient reflections from the residual. In this contribution, we use a sampling rate of \SI{48}{kHz} and set the block size to either 32~samples, for microphone arrays with 32 or fewer microphones, or to 64~samples, for arrays with more than 32~microphones. \revv{The hop size between consecutive blocks is set to $1/8$ of the block size to frequently update the thresholds, the residual estimate, and a possible decomposition.}

The detection threshold offset $\kappa$ determines the number of standard deviations by which the GSV sum in a signal block needs to exceed the averaged GSV sum such that a reflection is detected. With smaller values of $\kappa$, weaker energetic peaks are treated as reflections and with too small values most of the energy in the early part of the SRIR may be assigned to the direct part. If, on the other hand, $\kappa$ is chosen too large, only very strong reflections will be extracted from the residual. Values of $\kappa = 3$ or $4$ yielded good results in our experiments and are used in all examples in this contribution.

The averaging of the GSVs ensures that the proposed thresholds change smoothly over time and are not strongly influenced by individual reflections, cf.~Sec.~\ref{sec:thresholding}. If too little averaging is applied, the thresholds fluctuate strongly when reflections are detected and reflections that appear slightly earlier in time than other reflections might not be detected due to the raised detection threshold. (Recall that the algorithm proceeds backward in time.) If too much averaging is applied, the thresholds do not account for overall changes in energy in the residual. All examples in this work use averaging lengths between 32 and 64~blocks.

The length of the residual estimate determines how fast the GSV reacts to changes in the overall covariance of the residual. Appropriate lengths result in GSV sums that stay constant in the absence of reflections and exhibit strong peaks in the presence of reflections, cf.~Fig.~\ref{fig:thresholding}. Too short estimates prevent the implicit pre-whitening of the residual so that the GSV sums do not stay constant over time in the absence of reflections. Very long estimates reduce the relative peak height of GSV sums and thus make the separation between direct part and residual more difficult. All examples in this contribution use a residual estimate with a length of \SI{20}{ms}.

\subsection{Algorithm Summary}
This section summarizes the proposed algorithm with reference to the pseudocode in Algorithm~\ref{al:sub-dec}.

As illustrated in Fig.~\ref{fig:big-fig}~(a), the proposed algorithm starts by taking a signal block from the end of the SRIR as an initial residual estimate $\bm N_0$ and then advances in a blockwise manner toward the beginning of the SRIR, starting with the signal block $J$ that directly precedes the initial residual estimate. The FIFO buffer $\bm \rho$ that will later contain observations of the sum of the GSVs in the absence of salient reflections is initialized with very large values, or infinity, so that the detection threshold will not be exceeded within the first signal blocks. 

\rev{The signal is assumed to be divided into overlapping blocks before the processing and the GSVD is performed for each signal block $\bm X_i$ and the residual estimate $\bm N$.} The number of observations of the residual estimate $\bm N$ and the number of signal observations in the blocks $\bm X_i$ can be chosen independently. 
\rev{The exemplary SRIR from Figs.~\ref{fig:big-fig} and~\ref{fig:thresholding} was decomposed using a block size of ${K=32}$~samples, a hop size of 4~samples and a residual estimate with a length of 960~samples.}

The sum of the GSVs is then compared to the detection threshold that is calculated from the average of the observations of the GSV sum in $\bm \rho$ plus a multiple $\kappa$ of their standard deviation. \rev{For the exemplary SRIR, we set $\kappa = 4$ and averaged the GSV sum over 32~observations.}
If the current GSV sum exceeds the detection threshold, the cumulative sum of the GSVs is calculated, summing from the smallest toward the largest GSV. The estimated direct subspace dimension $Q_\s$ is obtained once the weighted cumulative GSV sum exceeds the average of the GSV sums in $\bm \rho$. From the dimension of the direct subspace, the binary direct subspace selection matrix $\bm \Gamma_\s$ is calculated. It is a diagonal matrix, containing ones as the first $Q_\s$ diagonal entries and zeros otherwise. 

If the current GSV sum does not exceed the detection threshold, $\bm \Gamma_\s$ is set to a zero matrix and the residual estimate $\bm N$ is updated \rev{by replacing its oldest rows by the rows of the current signal data matrix $\bm X_i$ that were not already included in previous signal blocks, using the FIFO principle.} Similarly, the current GSV sum replaces the oldest element in the vector $\bm \rho$.

Subsequently, the residual subspace selection matrix $\bm \Gamma_\n$ is calculated from the direct subspace selection matrix $\bm \Gamma_\s$. If the detection threshold was not exceeded, $\bm \Gamma_\n$ is the identity matrix, attributing all GSVs to the residual subspace. Otherwise, $\bm \Gamma_\n$ is a diagonal matrix, whose first $Q_\s$ diagonal entries are zero and whose last $Q_\n$ diagonal entries are one. Finally, the current signal block is decomposed into a direct part $\bm X_{\s,i}$ and a residual part $\bm X_{\n,i}$, by performing low-rank approximations of the signal matrix $\bm X_i$.

\begin{algorithm}[t]
\caption{SRIR Subspace Decomposition\label{al:sub-dec}}
\begin{algorithmic}[1]
\State $\bm N = \bm N_0$ \Comment{initial residual estimate}
\State $i=J$ \Comment{initial block index}
\State $\bm \rho = \textbf{inf}$ \Comment{initialize GSV sums as infinity}
\While{$i>1$} \Comment{iterate over signal blocks}
    \State $\bm X_i = \bm V_\x \bm \Sigma_\x \bm \Phi\T$ \Comment{GSVD of signal and residual data}
    \State $\bm N = \bm V_\n \bm \Sigma_\n \bm \Phi\T$
	\State $\bm \sigma = \dg{\bm \Sigma_\x\T \bm \Sigma_\x (\bm \Sigma_\n\T \bm \Sigma_\n)^{-1}}$ \Comment{GSVs}
	\State $\xi = \sum_{m=1}^M [\bm \sigma]_m$ \Comment{sum of GSVs}
	\If{$\xi > \text{mean}(\bm \rho) + \kappa\, \text{std}(\bm \rho)$} \Comment{reflection detected}
	    \State $c_\sigma = [\bm \sigma]_M$
	    \State $k = 1$
	    \While{$c_\sigma\, M/k < \text{mean}(\bm \rho)$}
	        \State $c_\sigma$\,+=\,$[\bm \sigma]_{M-k}$ \Comment{cumulative GSV sum}
	        \State $k$\,+\,+
	    \EndWhile
	    \State $Q_\s = M-k+1$ \Comment{direct subspace dimension}
	    \State $\bm \Gamma_\s = \bm I_{Q\times M}\T \bm I_{Q\times M}$
	\Else \Comment{no reflection detected}
	    \State $\bm \Gamma_\s = \bm 0_{M\times M}$ 
	    \State $\bm N \circlearrowleft \bm X_i$ \Comment{update residual estim. (FIFO)}
	    \State $\bm \rho \circlearrowleft \xi$ \Comment{update sum of GSVs (FIFO)}
    \EndIf
    \State $\bm \Gamma_\n = \bm I_{M\times M} - \bm \Gamma_\s$
    \State $\bm X_{\s,i} = \bm V_\x \bm \Sigma_\x \bm \Gamma_\s \bm \Phi\T$ \Comment{direct signal block}
    \State $\bm X_{\n,i} = \bm V_\x \bm \Sigma_\x \bm \Gamma_\n \bm \Phi\T$ \Comment{residual signal block}
	\State $i$\,-\,-
\EndWhile 
\end{algorithmic}
\end{algorithm}

\section{Quantitative Evaluation}\label{sec:comp_to_spat_sub}
This section comprises an evaluation of the proposed subspace decomposition method using simulated SRIRs. A perceptual evaluation of the method is beyond the scope of this contribution, however, a perceptual evaluation of an application of the herein proposed method is available in~\cite{Deppisch2022b}. The evaluation starts in Sec.~\ref{sec:space_vs_sh_domain} with an illustration showing that the application of the subspace decomposition is possible with unprocessed microphone array signals as well as with an SH decomposition thereof. The following sections apply the method to SH-domain signals to make it directly comparable to the spatial subtraction method whose signal model relies on SH-domain processing. In Sec.~\ref{sec:spectrum_analysis} magnitude spectra of extracted reflections that are obtained by the proposed method and the spatial subtraction method using two different signal models are analyzed. Then, in Sec.~\ref{sec:simulation_study}, the proposed method is evaluated using a spatio-spectral error measure and is compared to the spatial subtraction method and to a temporal cut-out approach for different rooms, microphone arrays, and levels of the residual. The evaluation ends with a comparison of the performance of the methods in the presence of two simultaneous reflections in Sec.~\ref{sec:simult_reflections}.
A case study with measured SRIRs follows in Sec.~\ref{sec:case_study_measured_srirs}.

\subsection{Raw vs. SH-Domain Processing}\label{sec:space_vs_sh_domain}
In a nutshell, the proposed subspace decomposition method achieves the separation of the direct part and the residual by comparing the energy of singular values of the array signals to the energy of a residual estimate. The method does not assume a particular arrangement or directivity of the employed microphones. An SH decomposition of the microphone signals can be interpreted as signals captured by microphones with a specific directivity, e.g., the zeroth-order SH has an omnidirectional directivity and first-order SHs have figure-of-eight directivities that are aligned with the Cartesian axes. Thus, the subspace decomposition method can be applied to unprocessed microphone signals or an SH decomposition thereof. 

\begin{figure}
\centering
\subfloat[]{\includegraphics[width=\columnwidth]{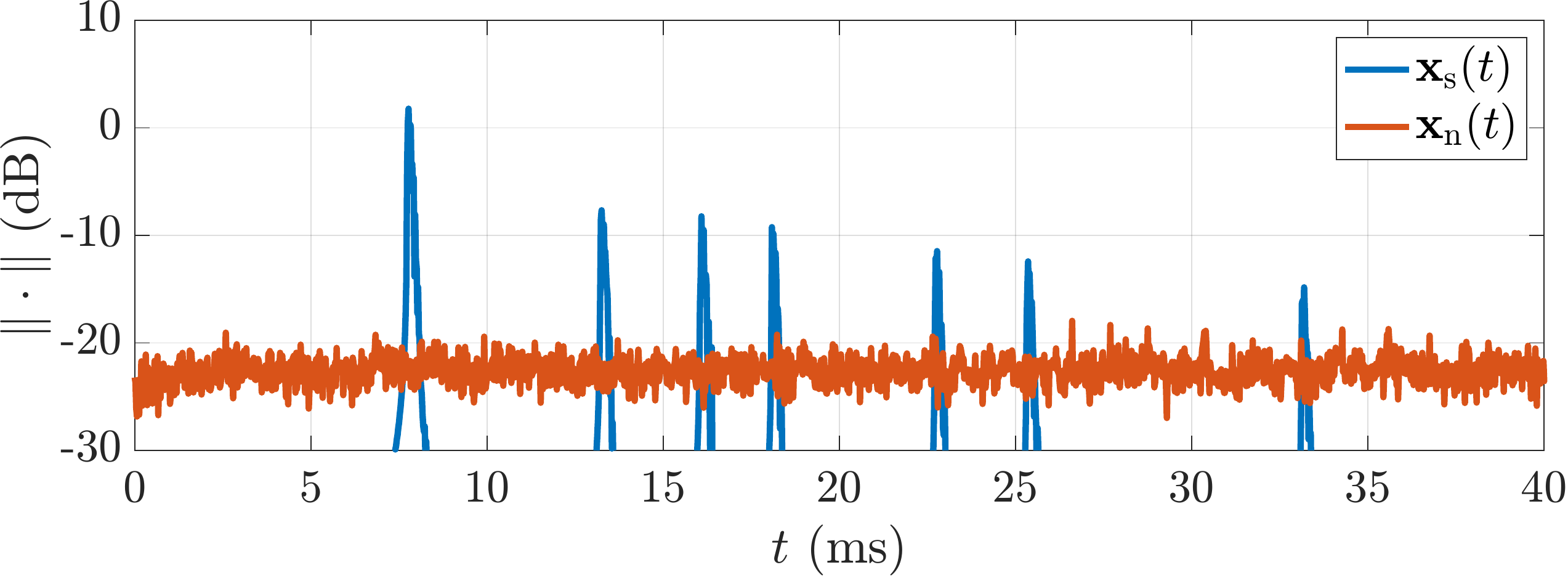}}\\
\vspace{-0.4cm}
\subfloat[]{\includegraphics[width=\columnwidth]{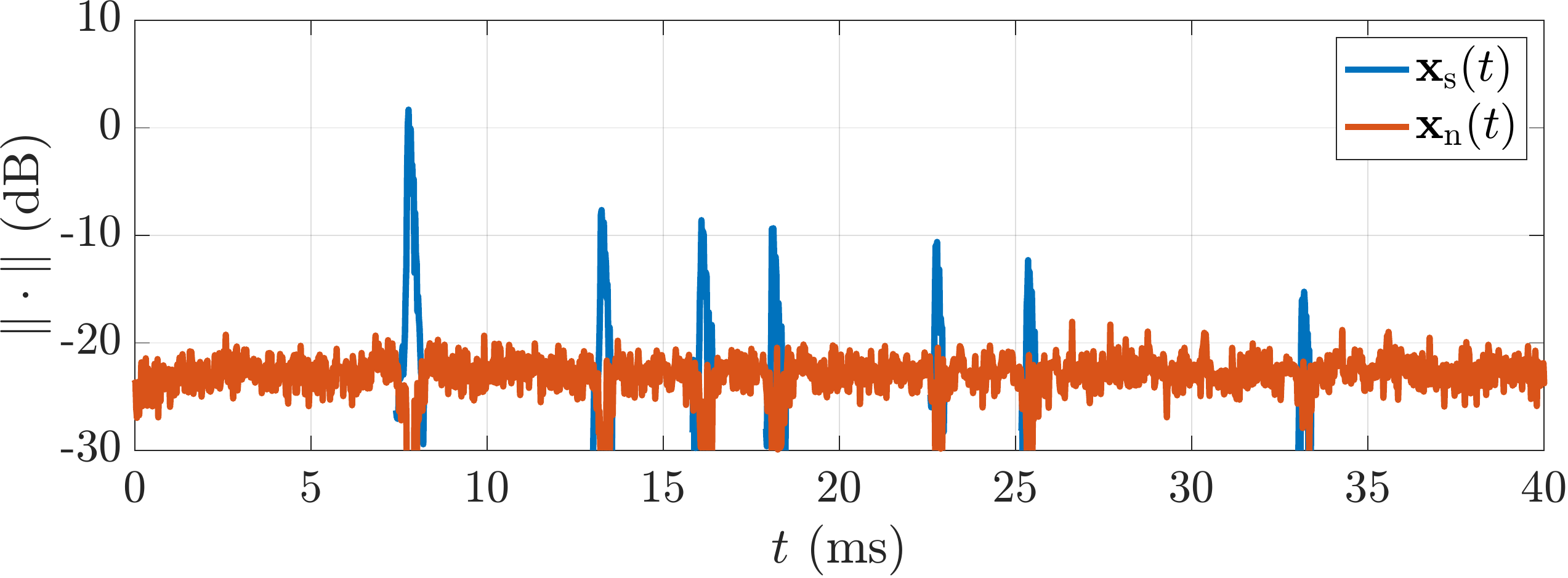}}\\
\vspace{-0.4cm}
\subfloat[]{\includegraphics[width=\columnwidth]{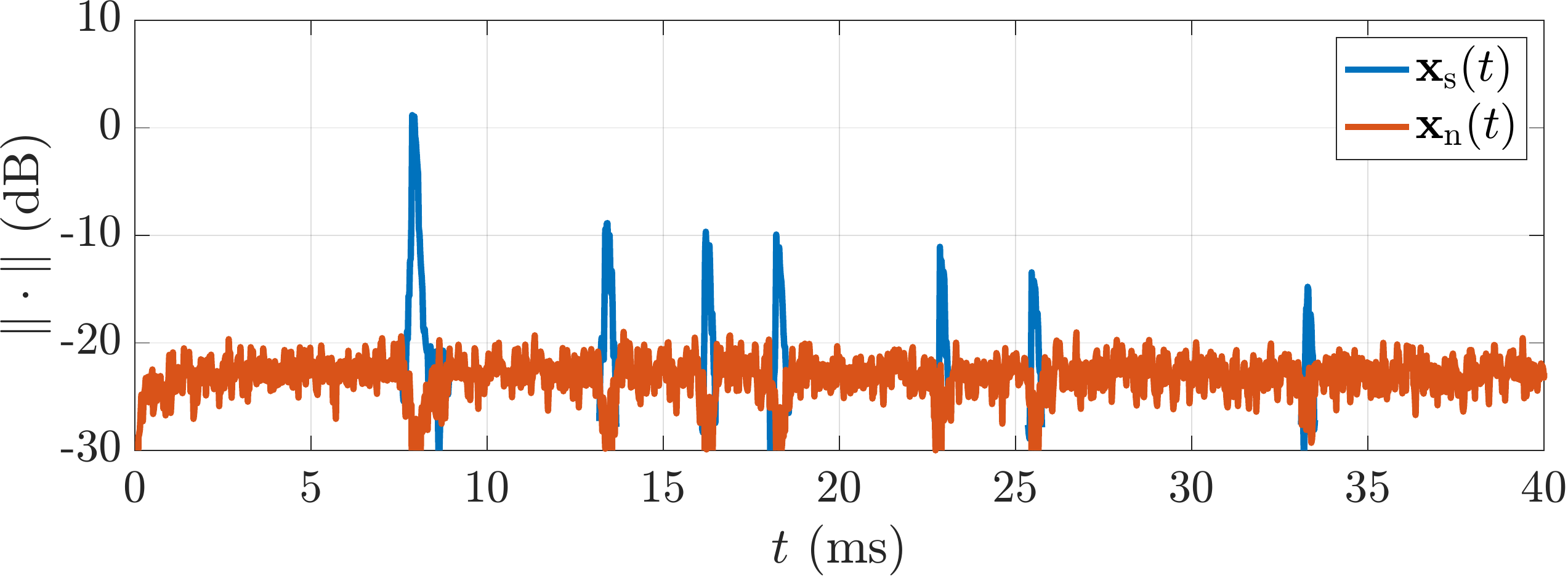}}
\vspace{-0.2cm}
\caption{Norms of the direct part $\bm x_\s(t)$ and the residual $\bm x_\n(t)$ of, \textbf{(a)}, the ground truth, \textbf{(b)}, the proposed method applied to unprocessed microphone signals and, \textbf{(c)}, the proposed method applied to an SH decomposition of the array signals.}
\label{fig:spaceVsShDomain}
\end{figure}
Fig.~\ref{fig:spaceVsShDomain}~(a) shows the norms of the ground truth direct part $\| \bm x_\s(t) \|$ and of the ground truth residual $\| \bm x_\n(t) \|$ of a simulated SRIR, i.e., seven simulated reflections are treated as direct part ground truth and noise with the spatial coherence of the array in an isotropic spherical noise field is treated as residual ground truth. The simulated, rigid array is again of radius \SI{4.2}{cm} and comprises 32~microphones that are arranged according to a t-design. 
Figs.~\ref{fig:spaceVsShDomain}~(b) and~(c) show the norms of the direct part $\| \bm x_\s(t) \|$ and of the residual $\| \bm x_\n(t) \|$ that are obtained by applying the subspace decomposition to the unprocessed SRIR and to an SH decomposition using up to fourth-order SHs. The subspace decomposition method does not have access to the individual parts shown in Fig.~\ref{fig:spaceVsShDomain}~(a) but only to their sum. All SH decompositions in this work are accompanied by radial filtering using Tikhonov regularization~\cite{Moreau2006}. The radial filtering reduces the influence of scattering on the SH signals and may improve the separability of reflections and the residual in more complex scenarios. A detailed analysis of this is however beyond the scope of this contribution and is left for future work. \revv{Following the reasoning from Sec.~\ref{sec:param_influence}}, the subspace decomposition was in both cases performed using a block size of 32~samples (\SI{0.7}{ms}), a hop size of 4~samples, a residual estimate of \SI{20}{ms}, GSV averaging of 32~blocks and $\kappa=4$.
A comparison of Figs.~\ref{fig:spaceVsShDomain}~(a), (b), and (c) shows that the decomposition is successful with unprocessed signals and with an SH decomposition thereof: in both cases, the seven reflections are extracted from the rest of the SRIR. A detailed performance evaluation using a numerical error measure follows in Sec.~\ref{sec:simulation_study}.

\subsection{Analysis of Extracted Reflection Spectra}\label{sec:spectrum_analysis}
In this section, the norms of the spectra $\bm \chi_\mathrm{s}(f)$ of two of the ground truth reflections from Fig.~\ref{fig:spaceVsShDomain}~(a) are compared to extracted spectra from the direct part $\bm x_\s(t)$ obtained either via the spatial subtraction method~\cite{Deppisch2021b} or the subspace decomposition. The frequency-domain vector $\bm \chi_\mathrm{s}(f)$ contains the spectrum of all SH-domain signal channels during the presence of a reflection. Specifically, we analyze the spectra of the first and the last reflection in Fig.~\ref{fig:spaceVsShDomain}~(a) to show results for different ratios of reflection and residual energy. According to Parseval's theorem for the spherical Fourier transform~\cite{Rafaely2007}, the norm of an SH-domain signal vector equals the total signal energy integrated over the surface of the unit sphere and is thus a suitable measure to illustrate the overall results of the different methods. For the spatial subtraction method, we assume that the time-of-arrival (TOA) of the respective reflection is known and use SH-MUSIC~\cite{Khaykin2009} to estimate its DOA. For both reflections, the DOA estimation errors are small, they amount to $1.6^\circ$ for the first reflection and to $2.1^\circ$ for the seventh reflection. After the DOA estimation, the spatial subtraction method is applied using two different signal models, the one originally proposed in the context of sound scenes in~\cite{Politis2018}, in the following referred to as \textit{SpatSub1}, and the comprehensive signal model from~\cite{Deppisch2021b} that includes the influences of scattering, radial filtering, and spatial aliasing and is referred to as \textit{SpatSub2}. Note that the proposed subspace decomposition method, which is in the following also referred to as \textit{SubDec}, does not have access to the true TOAs and does not utilize the DOA estimates. The spatial subtraction method was applied using the discrete Fourier transform (DFT) of the array signals within a \SI{1}{ms} rectangular window centered around the respective reflection and the shown spectra are calculated within the same window. \revv{The \SI{1}{ms} window ensures that the full reflection is captured while limiting the amount of noise, i.e., the norm of the simulated reflection decays by about \SI{30}{dB} within the window and is at least \SI{6}{dB} below the noise floor at the edge of the window.} The subspace decomposition was performed using the same parameters as in Sec.~\ref{sec:space_vs_sh_domain}.

\begin{figure}
\centering
\subfloat[]{\includegraphics[width=\columnwidth]{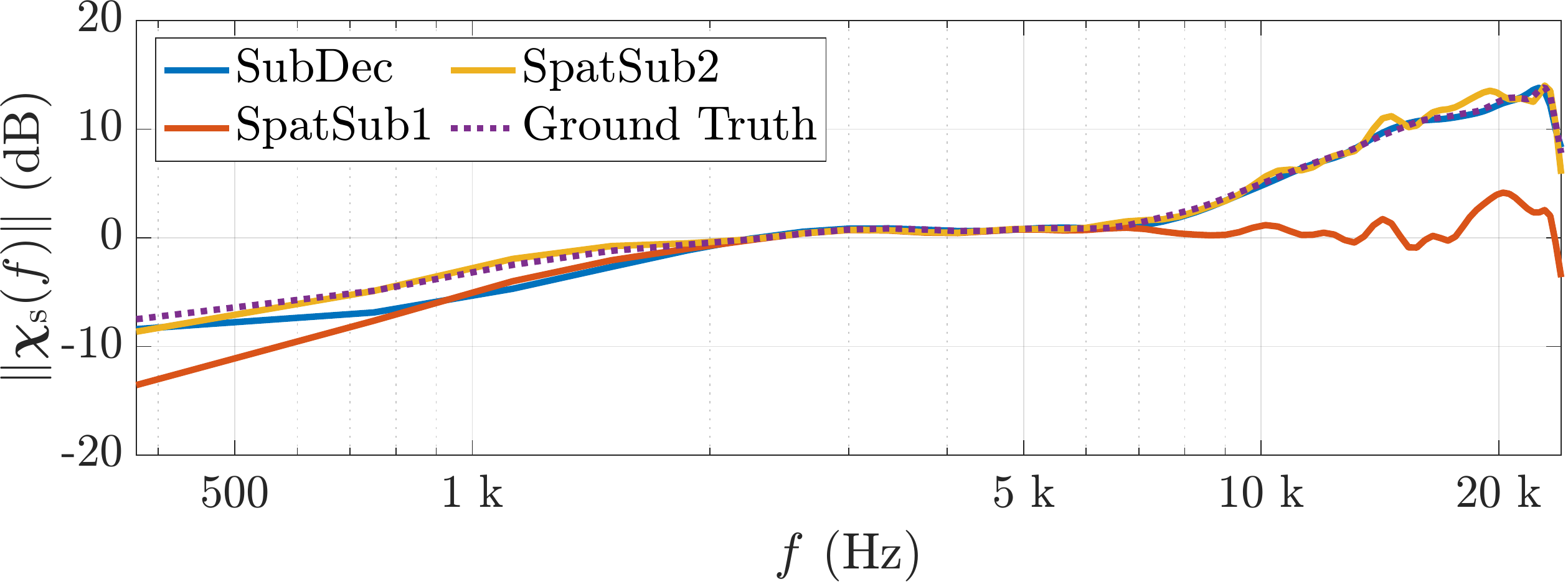}}
\\
\vspace{-0.4cm}
\subfloat[]{\includegraphics[width=\columnwidth]{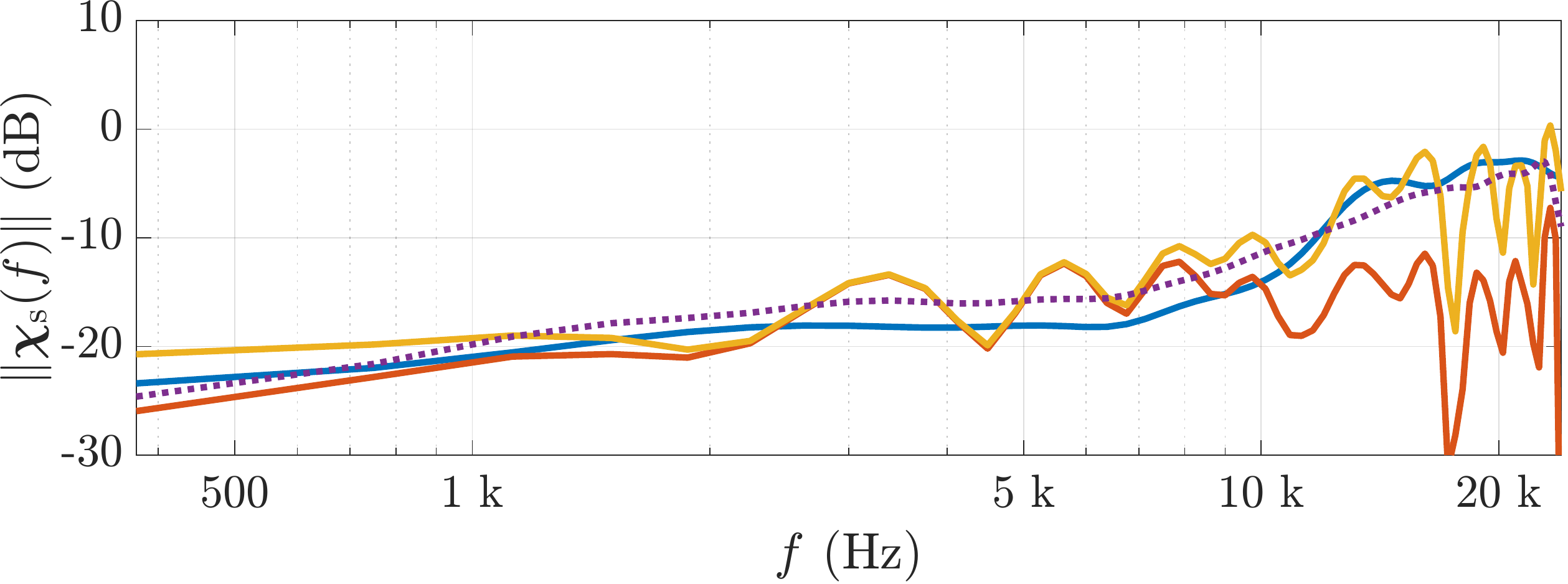}}
\vspace{-0.2cm}
\caption{Norms of the ground truth spectra $\bm \chi_\mathrm{s}(f)$ of, \textbf{(a)}, the first and, \textbf{(b)}, the seventh reflection from Fig.~\ref{fig:spaceVsShDomain}~(a) and of extracted reflection spectra using the spatial subtraction method with two different signal models, \textit{SpatSub1} and \textit{SpatSub2}, as well as the proposed subspace decomposition method \textit{SubDec}.}
\label{fig:extractedSpectra}
\end{figure}
Fig.~\ref{fig:extractedSpectra}~(a) shows the obtained norms of the spectra $\bm \chi_\mathrm{s}(f)$ for the first reflection. The norms of the spectra obtained by both \textit{SubDec} and \textit{SpatSub2} follow the norm of the ground truth closely. The subspace decomposition (\textit{SubDec}) shows a maximum deviation from the ground truth of about \SI{2}{dB} around \SI{1}{kHz} while \textit{SpatSub2} has a maximum deviation of about \SI{1.5}{dB} around \SI{19}{kHz}. The method \textit{SpatSub1} deviates more strongly from the ground truth. Its underlying signal model does not include the influence of non-ideal radial filtering, leading to a deviation of about \SI{6}{dB} around \SI{400}{Hz}, and also neglects the influence of spatial aliasing and the SH order truncation, leading to large deviations above \SI{8}{kHz}, with a maximum deviation of \SI{12}{dB} around \SI{16}{kHz}.

The norms of the spectra for the extraction of the seventh reflection are shown in Fig.~\ref{fig:extractedSpectra}~(b). The ratio of reflection peak to residual energy is in this case much lower in comparison to the first reflection, cf.~Fig.~\ref{fig:spaceVsShDomain}~(a), making the extraction task more difficult. Again \textit{SubDec} and \textit{SpatSub2} follow the spectrum of the ground truth closely but this time \textit{SpatSub2} exhibits strong fluctuations that increase with frequency, resulting in a maximum deviation of \SI{13}{dB} at \SI{17}{kHz}. The spectrum of the proposed method \textit{SubDec} does not show such fluctuations and has a maximum deviation from the ground truth of about \SI{2.6}{dB} around \SI{6}{kHz}. The method \textit{SpatSub1} shows similar fluctuations as \textit{SpatSub2} but additionally deviates strongly from the ground truth at high frequencies above \SI{10}{kHz}.

\subsection{Simulation Study}\label{sec:simulation_study}
\begin{figure*}
\centering
\subfloat[]{\includegraphics[width=\columnwidth]{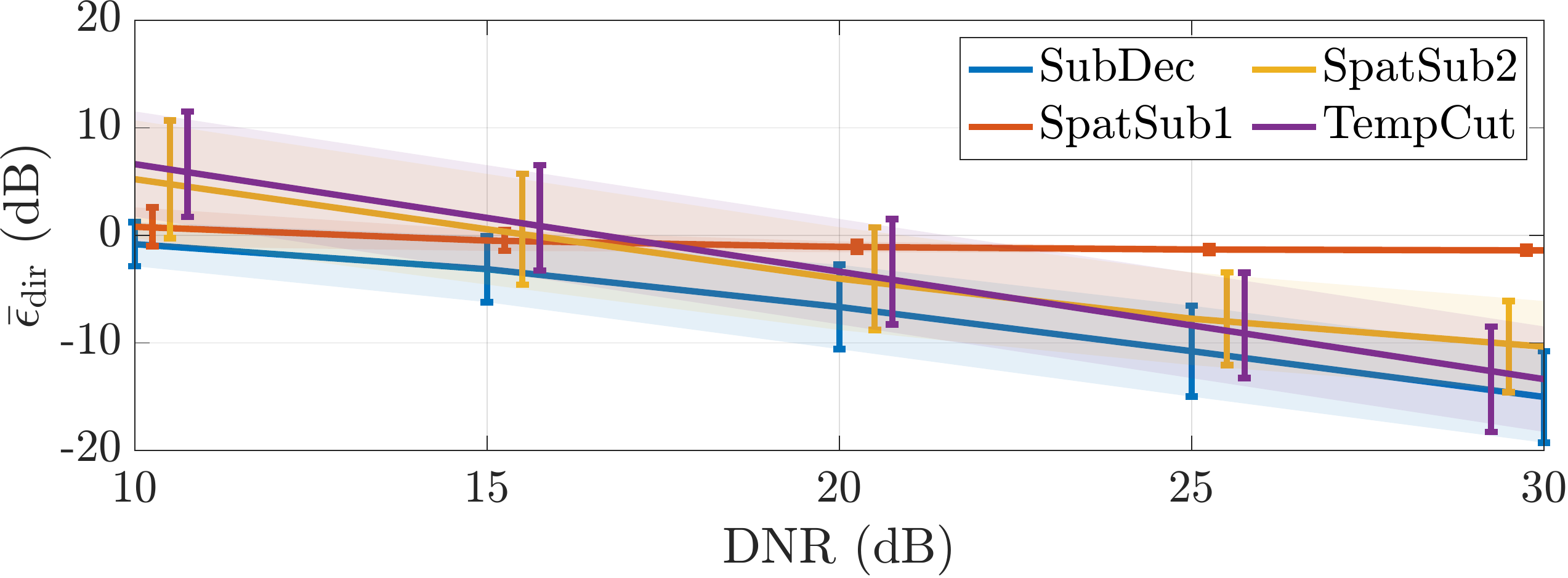}}
\hfill
\subfloat[]{\includegraphics[width=\columnwidth]{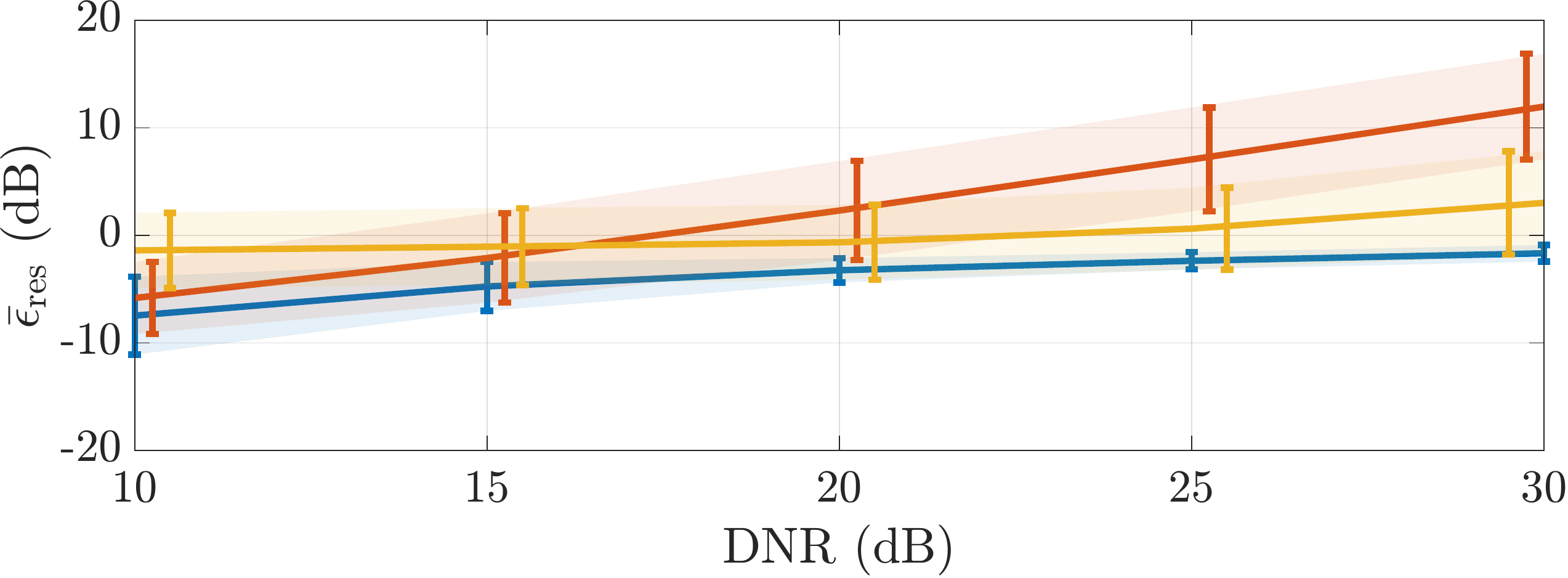}}\\
\vspace{-0.4cm}
\subfloat[]{\includegraphics[width=\columnwidth]{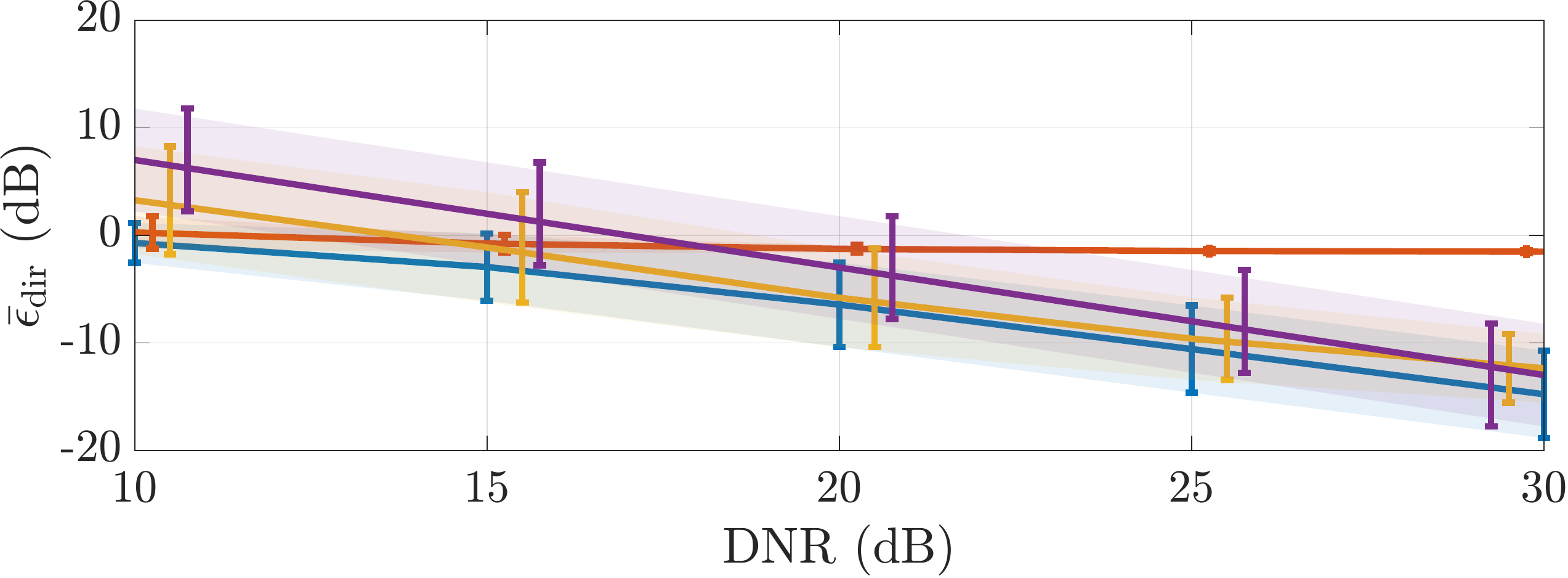}}
\hfill
\subfloat[]{\includegraphics[width=\columnwidth]{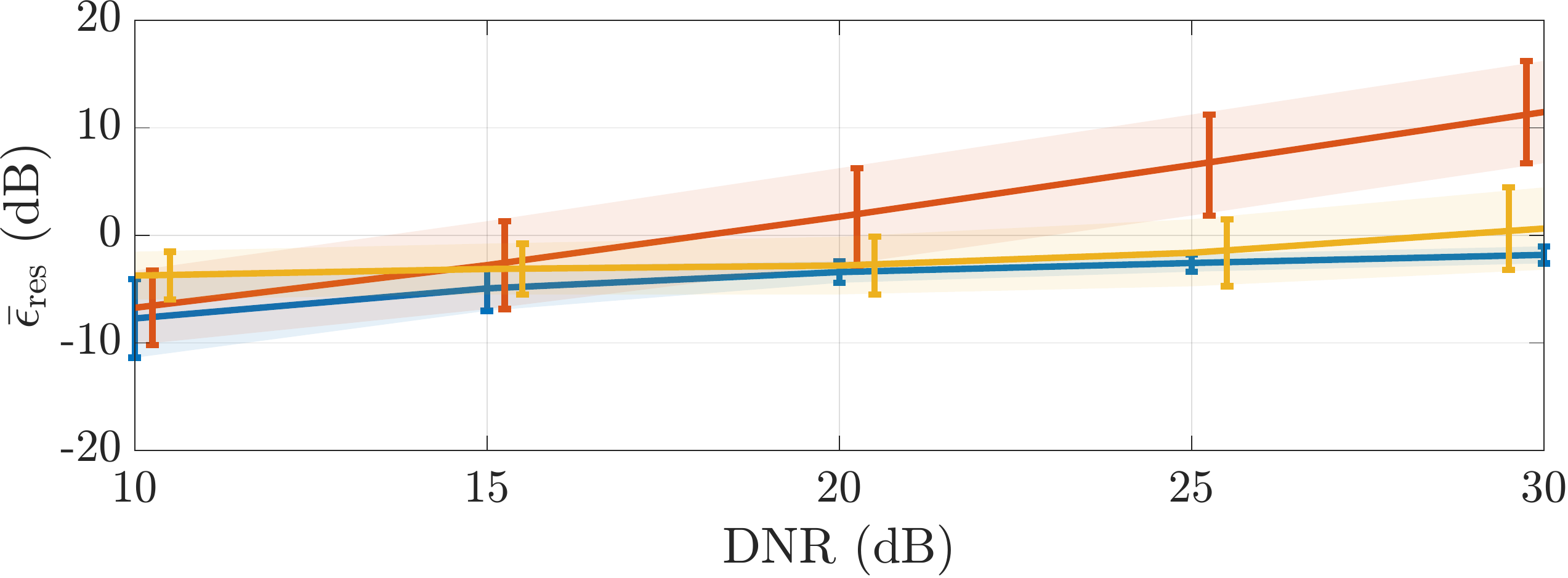}}\\
\vspace{-0.4cm}
\subfloat[]{\includegraphics[width=\columnwidth]{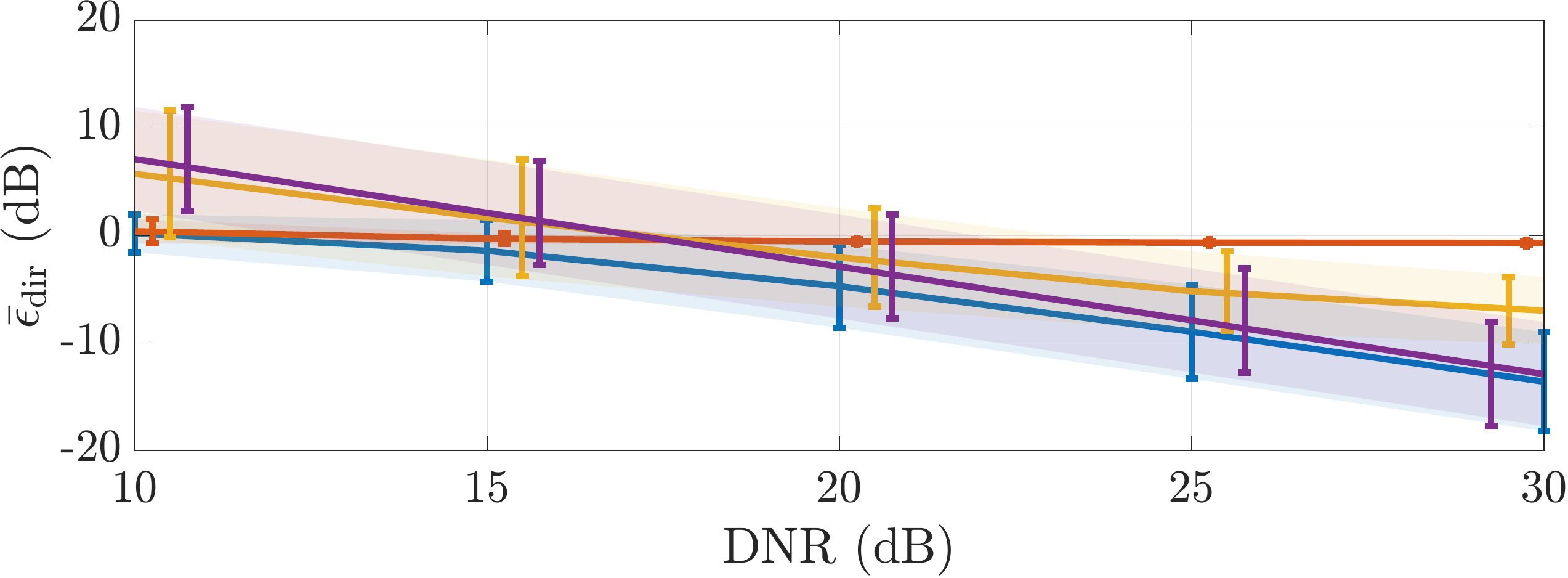}}
\hfill
\subfloat[]{\includegraphics[width=\columnwidth]{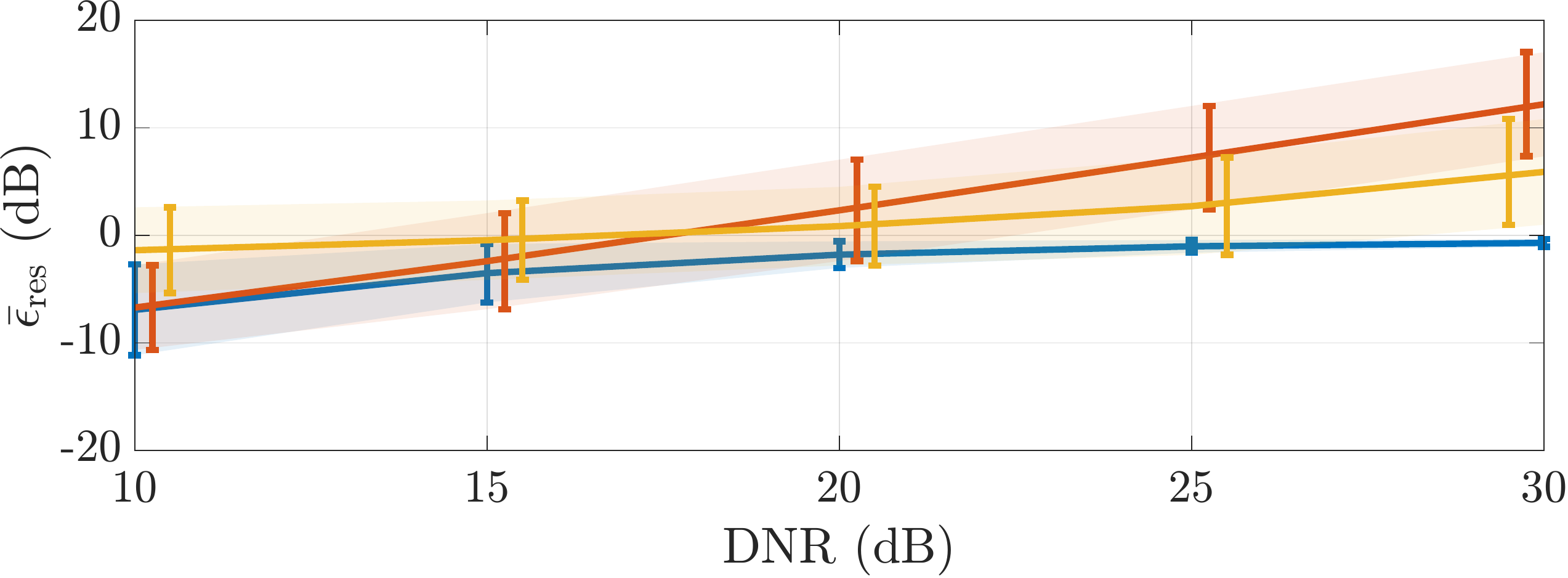}}
\vspace{-0.2cm}
\caption{Means and standard deviations of the average direct part error $\bar{\epsilon}_\mathrm{dir}$ (left column) and the average residual error $\bar{\epsilon}_\mathrm{res}$ (right column) of the compared methods \revv{as a function of the omnidirectional direct-sound-peak-to-residual-noise-RMS ratio (DNR)}, for, \textbf{(a)} and \textbf{(b)}, \textit{Array1} with 24~microphones, \textbf{(c)} and \textbf{(d)}, \textit{Array2} with 32~microphones, and, \textbf{(e)} and \textbf{(f)}, \textit{Array3} with 48~microphones.}
\label{fig:simStudy}
\end{figure*}

To systematically evaluate the performance of the proposed method, a simulation study is performed. SMIRGen was again used to simulate the direct part of SRIRs as first-order image sources. No higher-order image sources were calculated to be able to assume that all generated reflections can be considered salient and therefore be assigned to the direct part. This was further achieved by setting the broadband absorption coefficient to 0.3 in all simulations. The simulation study comprises the combination of, (i), 15~shoebox-shaped rooms with random dimensions, (ii), 3~different spherical microphone arrays, (iii) the direct sound and six first-order reflections per SRIR, and, (iv), five different ratios of direct-sound-peak to residual root-mean-square (RMS) energy. The rooms were generated with uniformly-distributed dimensions between $4 \times 4 \times 2$~\si{m} and $15 \times 15 \times 10$~\si{m}. The source and microphone array positions were randomly generated with the constraints of having a distance of at least \SI{1}{m} to any room boundary and at least \SI{2}{m} from each other. Additionally, it was ensured that the generated reflections arrive at the microphone array with a time difference of at least \SI{1}{ms} so that for the subspace decomposition methods it can be assumed that each subtraction window contains a single reflection. The case of two simultaneously arriving reflections will be investigated in Sec.~\ref{sec:simult_reflections}. The residual part was generated as noise with the coherence of the simulated arrays in an isotropic diffuse field with a decay of \SI{60}{dB} per second. An SH decomposition was performed for both the direct part and the residual, and both parts were radial filtered before being added together with varying energy ratios. For this purpose, we define the omnidirectional direct-sound-peak-to-residual-noise-RMS ratio (DNR) that comprises the ratio between the maximum absolute value of the zeroth-order SH channel and the zeroth-order-SH RMS value of the generated residual noise. Note that we define the DNR as a measure per SRIR, meaning that the direct sound of the simulated SRIR is ensured to stand out against the residual RMS but this is not necessarily the case for the six first-order reflections. The DNR was varied in \SI{5}{dB} steps between \SI{10}{dB} and \SI{30}{dB}. 
The three microphone arrays under test are all rigid, spherical arrays of radii \SI{4.2}{cm}, \SI{4.2}{cm}, and \SI{8.5}{cm}. They comprise $24$, $32$, and $48$~microphones that are arranged according to t-designs and allow for SH decompositions of maximum order $3$, $4$, and $5$. They will also be referred to as \textit{Array1}, \textit{Array2}, and \textit{Array3} in the following.

\begin{figure}
    \centering
    \includegraphics[width=\columnwidth]{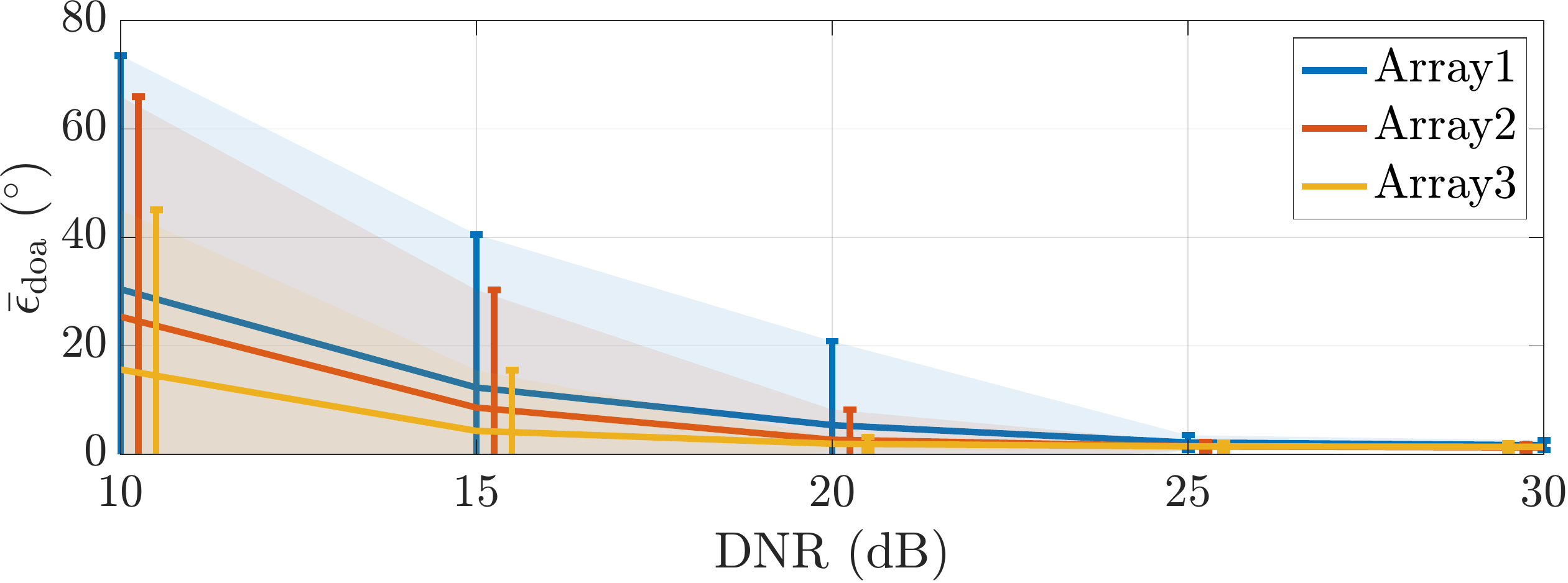}
    \caption{Mean and standard deviation of the average DOA estimation error $\bar{\epsilon}_\mathrm{doa}$ for the three microphone arrays and different DNRs.}
    \vspace{-0.4cm}
    \label{fig:doaErrorSimStudy}
\end{figure}
To facilitate a numerical evaluation, we define a spatio-spectral error measure that comprises the ratio of the norm of the difference of the spectrum of the ground truth reflection $\bm \chi_\mathrm{s}^\mathrm{gt}(f_b)$ and the spectrum of an extracted reflection $\bm \chi_\mathrm{s}(f_b)$, to the norm of the ground truth reflection,
\begin{equation}\label{eq:error_measure}
    \epsilon_\mathrm{dir} = \frac{\sum_b \| \bm \chi_\mathrm{s}(f_b)-\bm \chi_\mathrm{s}^\mathrm{gt}(f_b) \|}{\sum_b \| \bm \chi_\mathrm{s}^\mathrm{gt}(f_b) \|} \, ,
\end{equation}
where the sum over $b$ denotes the sum over all frequency bins of a 128-point DFT.
The division by the norm of the ground truth spectrum ensures that the calculated error is relative to the ground truth energy, i.e., low-energy reflections do not automatically generate a lower error. The spatio-spectral residual error $\epsilon_\mathrm{res}$ is defined similarly by replacing the reflection spectrum $\bm \chi_\mathrm{s}(f_b)$ and its ground truth $\bm \chi_\mathrm{s}^\mathrm{gt}(f_b)$ by the spectrum of the residual $\bm \chi_\mathrm{n}(f_b)$ and its ground truth $\bm \chi_\mathrm{n}^\mathrm{gt}(f_b)$. All spectra are calculated within a \SI{1}{ms} window that is centered around the ground truth TOA of the reflection.

As before, the spatial subtraction methods have access to the true TOA of the reflections and use SH-MUSIC for the DOA estimation. The subspace decomposition method is applied without access to any additional information from the ground truth and does not require DOA estimation. It is performed using a block size of 32~samples for \textit{Array1} and \textit{Array2}, and 64~samples for \textit{Array3}, a hop size of $1/8$ of the block size, a residual estimate of \SI{20}{ms}, GSV averaging of 32~blocks and $\kappa=4$.
To increase the interpretability of the results, another approach is added to the comparison that involves a temporal cut-out of the reflections. It is similarly performed in~\cite{Puomio2021a} and, using an omnidirectional RIR, in~\cite{Porschmann:DAFx2017}. The approach is in the following also referred to as \textit{TempCut} and comprises cutting out individual reflections via a \SI{1}{ms} window that is centered around the reflection. The cut-out is equally applied to all SH channels and is a straightforward approach that avoids the need for beamforming or a subspace decomposition. However, the \textit{TempCut} approach cannot provide a residual SRIR and is thus only considered in the direct-part comparison of the methods.

Fig.~\ref{fig:simStudy} shows the mean and the standard deviation of the average direct part error $\bar{\epsilon}_\mathrm{dir}$ and the average residual error $\bar{\epsilon}_\mathrm{res}$, i.e., both errors are averaged over the different rooms and over the individual reflections. Errors are shown for the three different microphone arrays and for different DNRs. The proposed subspace decomposition method \textit{SubDec} outperforms the compared methods in all tested cases. The two spatial subtraction methods perform differently, depending on the DNR. For lower DNRs, \textit{SpatSub1} outperforms \textit{SpatSub2} while for higher DNRs \textit{SpatSub2} outperforms \textit{SpatSub1}. For lower DNRs, the residual noise prevents an accurate DOA estimate and distorts the estimate of the reflection spectrum of the underlying plane-wave signal model. These errors more severely influence the results from \textit{SpatSub2} as small inaccuracies have a big influence on the estimate of the spatial aliasing. The mean and standard deviation of the average DOA estimation error $\bar{\epsilon}_\mathrm{doa}$ are shown in Fig.~\ref{fig:doaErrorSimStudy}.

Two overall trends can be observed for all methods: the average direct part errors $\bar{\epsilon}_\mathrm{dir}$ tend to decrease with increasing DNR and the average residual errors $\bar{\epsilon}_\mathrm{res}$ tend to increase with increasing DNR. With increasing DNR, individual reflections stand out more against the residual and are hence easier to extract. Further, as shown in Fig.~\ref{fig:doaErrorSimStudy}, the DOA estimates that are required for the spatial subtraction methods get more accurate with higher DNR. For high DNRs, $\bar{\epsilon}_\mathrm{dir}$ of the \textit{TempCut} method approaches the results of \textit{SubDec} because at high DNRs the residual energy is negligible in comparison to the direct part energy. At low DNRs, \textit{TempCut} performs worst as the residual dominates over the reflection and a simple temporal cut-out thus creates a large error. 

In the case of the average residual errors $\bar{\epsilon}_\mathrm{res}$, an increase in error can be observed with increasing DNR. Although the reflections stand out more against the residual with higher DNRs and thus the extraction task becomes simpler, the generated errors increase as the employed error measure, cf.~\eqref{eq:error_measure}, is normalized by the energy of the ground truth residual. At high DNRs, the ground truth residual carries low energy and thus relative errors tend to increase with the DNR.

\subsection{Two Simultaneous Reflections}\label{sec:simult_reflections}
\begin{figure}
\centering
\subfloat[]{\includegraphics[width=\columnwidth]{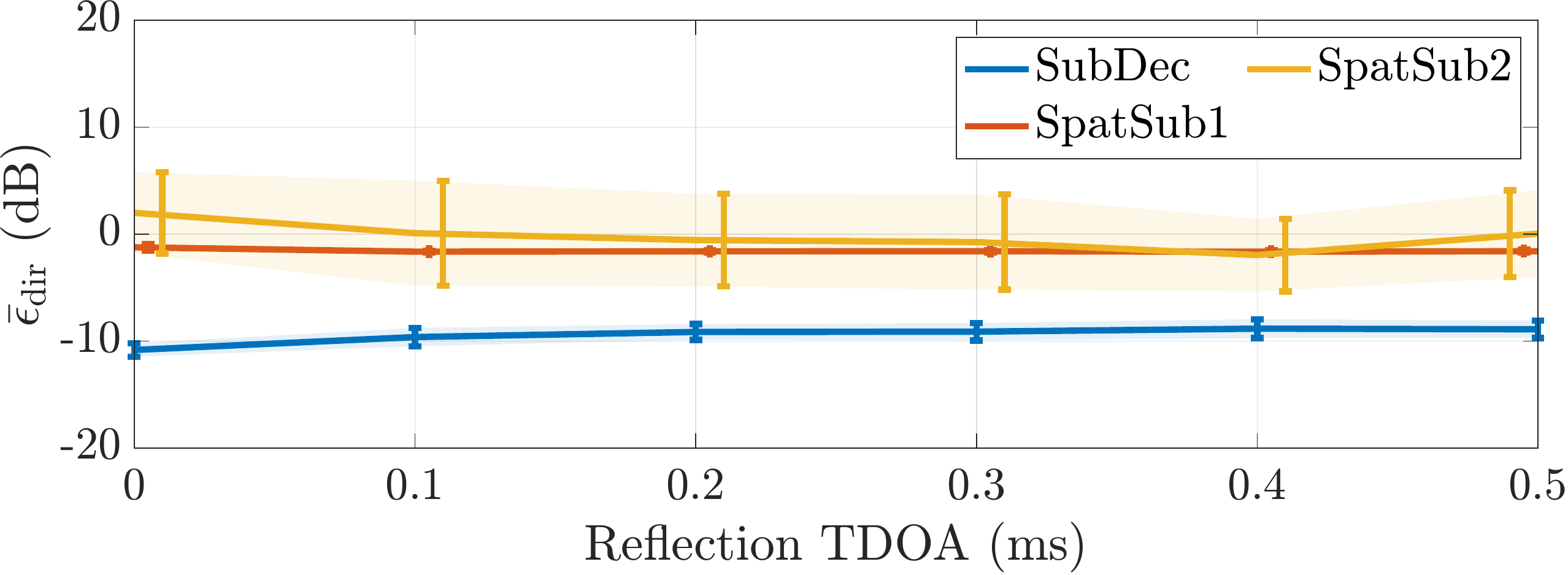}}
\\
\vspace{-0.4cm}
\subfloat[]{\includegraphics[width=\columnwidth]{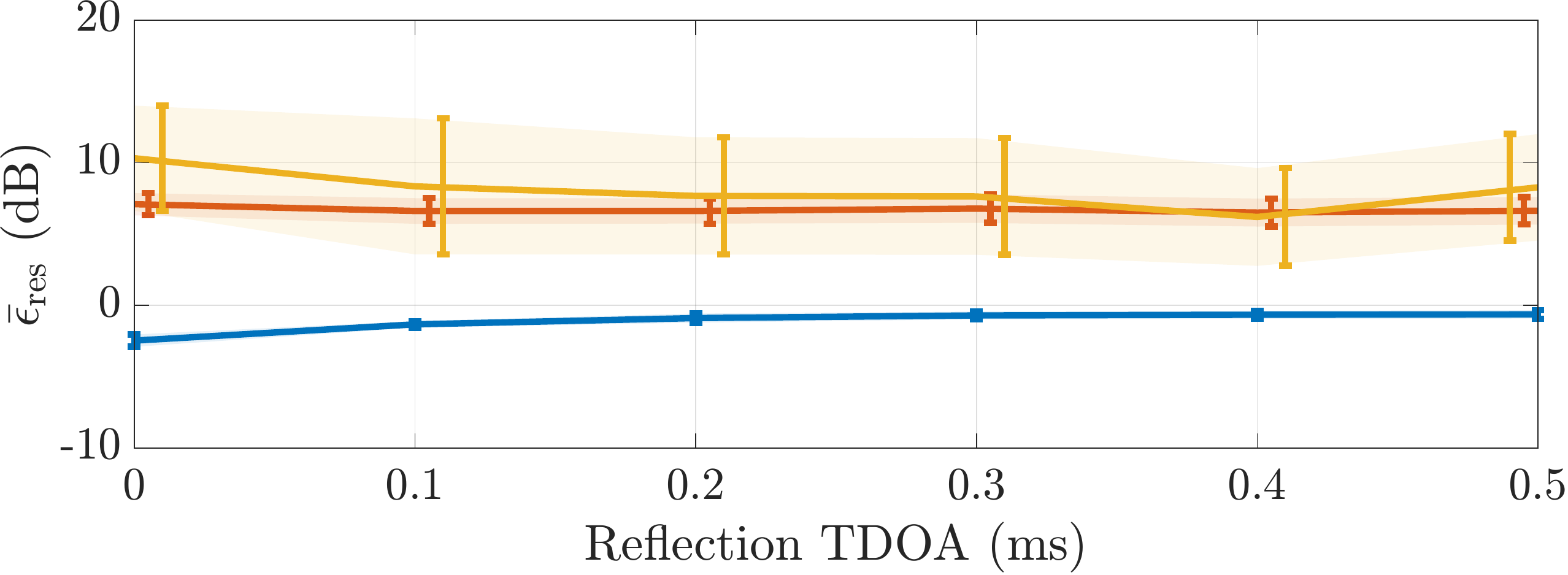}}
\vspace{-0.2cm}
\caption{Means and standard deviations of, \textbf{(a)}, the average direct part error $\bar{\epsilon}_\mathrm{dir}$ and, \textbf{(b)}, the average residual error $\bar{\epsilon}_\mathrm{res}$ for two simultaneous reflections with different TDOAs. The simulated array is \textit{Array2}.}
\label{fig:simultRefl}
\end{figure}
The reflection density in acoustic environments typically increases exponentially with time. Thus, multiple reflections are likely to occur within one analysis signal block of the different decomposition algorithms. In the following, we refer to multiple reflections within one signal block as simultaneous reflections and investigate the decomposition performance of the different algorithms for two simultaneous reflections. While this does not require a modification of the subspace decomposition method, the spatial subtraction method using both signal models is extended as in~\cite{Politis2018} to support the simultaneous subtraction of multiple reflections. The number of simultaneous reflections and their average time of arrival is assumed to be known by the spatial subtraction methods and the spatial subtraction window of \SI{1}{ms} length is centered around the average TOA of the two reflections. We perform simulations for time-differences-of-arrival (TDOAs) of the reflections at the array center between \SI{0}{ms} and \SI{0.5}{ms}. Both reflections are created with the same magnitude and different angles of arrival. For each simulated TDOA, 100~repetitions with random, unique incidence angles that are drawn from the vertices of a dodecahedron are performed. This ensures that the two reflections have at least an angular separation of $41^\circ$. The subspace decomposition is performed using the same parameters as in Sec.~\ref{sec:simulation_study}. The simulated rigid, spherical array is the \textit{Array2} from the previous simulation, i.e., its 32~microphones are distributed according to a t-design and a fourth-order SH decomposition is performed. Non-decaying noise with the coherence of the array in an isotropic diffuse field is added to achieve a DNR of \SI{20}{dB}.

\begin{figure}
    \centering
    \includegraphics[width=\columnwidth]{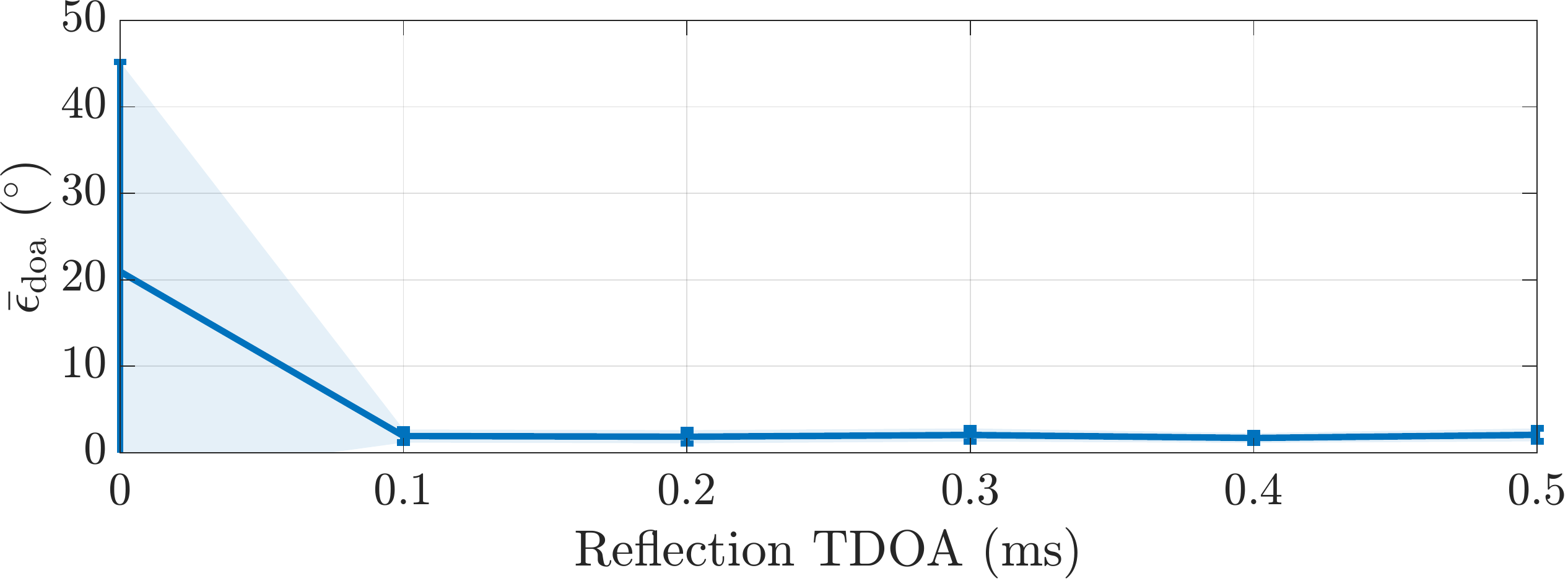}
    \vspace{-0.4cm}
    \caption{Means and standard deviations of the average DOA estimation error $\bar{\epsilon}_\mathrm{doa}$ for two simultaneously arriving reflections, \textit{Array2} and different TDOAs.}
    \label{fig:simultReflDoaError}
\end{figure}
Fig.~\ref{fig:simultRefl} shows the means and the standard deviations of the average direct part errors $\bar{\epsilon}_\mathrm{dir}$ and the average residual errors $\bar{\epsilon}_\mathrm{res}$ for TDOAs between \SI{0}{ms} and \SI{0.5}{ms} and Fig.~\ref{fig:simultReflDoaError} shows the corresponding average DOA estimation errors $\bar{\epsilon}_\mathrm{doa}$. Again, the subspace decomposition method outperforms both spatial subtraction methods in terms of both direct part error and residual error for all TDOAs, although the average DOA estimation error means are equal to or below $2^\circ$ for TDOAs of \SI{0.1}{ms} or more. In contrast to the previous simulations of the array with a DNR of \SI{20}{dB}, cf.~Figs.~\ref{fig:simStudy}~(c) and~(d), \textit{SpatSub1} now achieves lower errors than \textit{SpatSub2} except for the case with a TDOA of \SI{0.4}{ms}. The plane-wave model parameters of the comprehensive signal model of \textit{SpatSub2} cannot be estimated accurately due to the interference of the two reflections, which creates the observed error.

\section{Case Study With Measured Spatial Room Impulse Responses}\label{sec:case_study_measured_srirs}
To demonstrate the practical applicability of the proposed method, we apply the subspace decomposition to three SRIRs that were measured in different acoustic environments. The three SRIRs cover a large variety of acoustic conditions and include some variation in terms of the employed microphone arrays. All three SRIRs are publicly available. \rev{Binaural renderings of the original SRIRs and the direct part and residual SRIRs from the subspace decomposition are provided on a companion website\footnote{\url{http://www.ta.chalmers.se/srir-subspace-decomposition/}}.}

\begin{figure}
\centering
\subfloat[]{\includegraphics[width=\columnwidth]{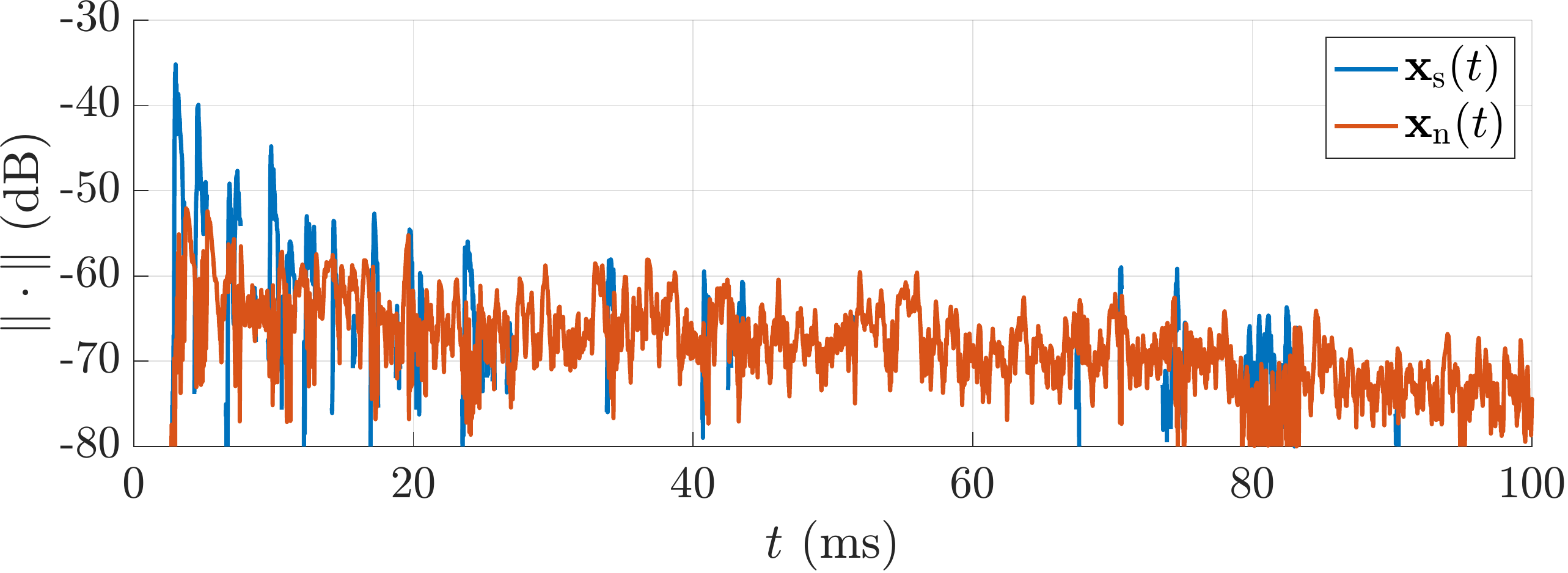}}
\\
\vspace{-0.4cm}
\subfloat[]{\includegraphics[width=\columnwidth]{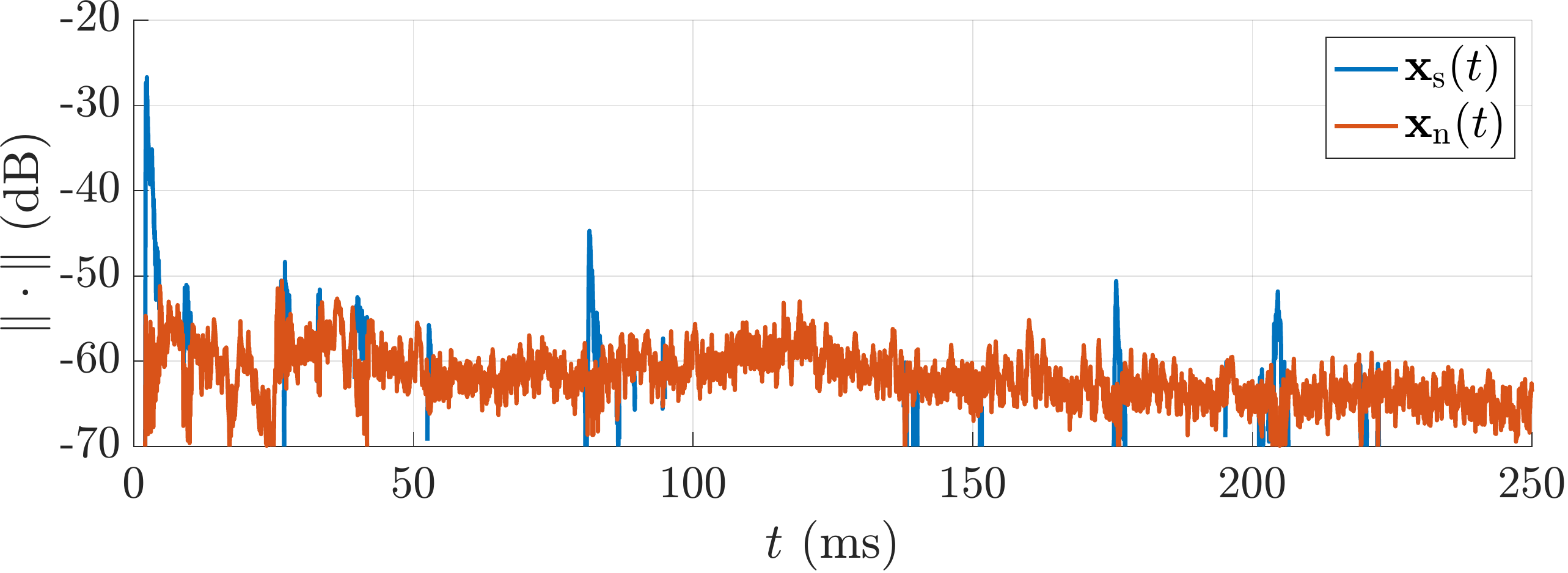}}
\\
\vspace{-0.4cm}
\subfloat[]{\includegraphics[width=\columnwidth]{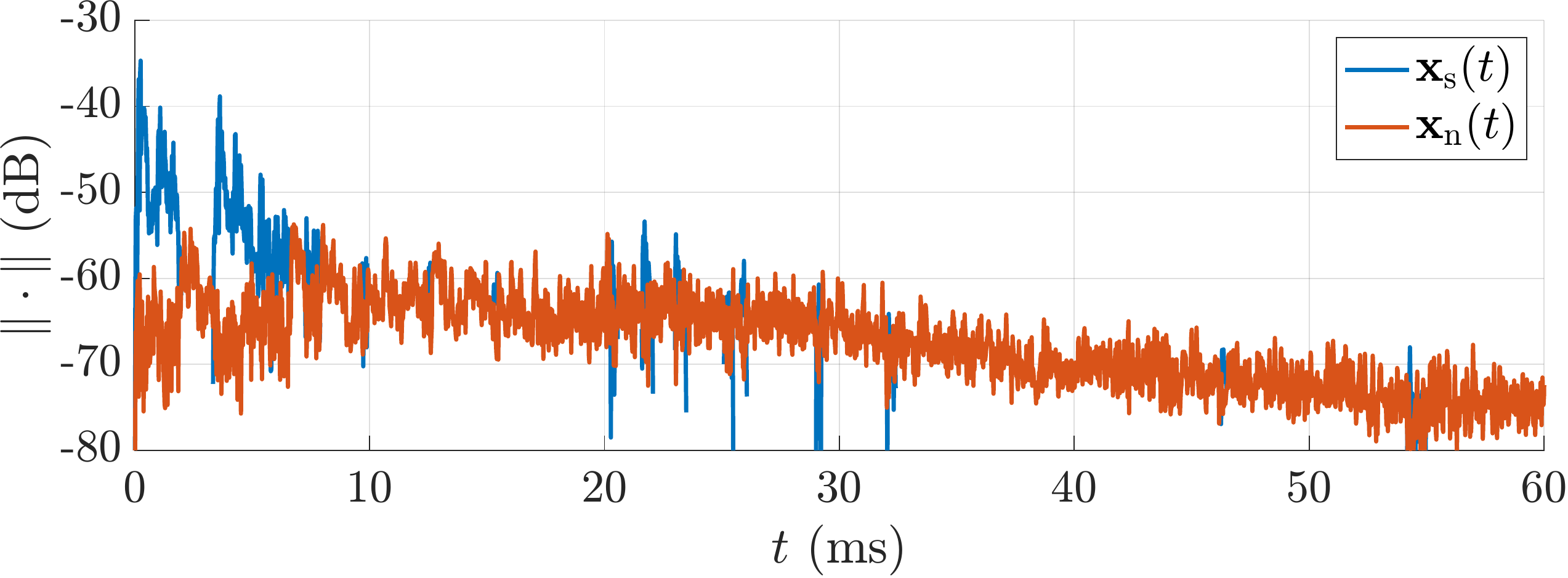}}
\vspace{-0.2cm}
\caption{Norms of the direct part $\bm x_\s(t)$ and the residual $\bm x_\n(t)$ of SRIRs measured, \textbf{(a)}, in a small conference room, \textbf{(b)}, in a concert hall, and, \textbf{(c)}, in the entrance of an anechoic chamber with the measurement loudspeaker located in the adjacent office.}\label{fig:caseStudy}
\end{figure}

The first SRIR was measured in a $10.3 \times 5.8 \times 3.1$~\si{m} conference room with a broadband reverberation time of \SI{0.63}{s}~\cite{Schneiderwind2019}. The measurement was performed using the Eigenmike em32 32-channel rigid-sphere microphone array with a radius of \SI{4.2}{cm}. \rev{The subspace decomposition was performed using a block size of 32~samples (\SI{0.7}{ms}) and a hop size of 4~samples. The residual estimate had a length of \SI{20}{ms} and the thresholds were calculated with a GSV averaging length of 64~blocks and using $\kappa = 3$. Fig.~\ref{fig:caseStudy}~(a) shows the direct and residual subspace decomposition for measurement position~2 of the data set.}

The second SRIR was measured in a $30.3 \times 16.5 \times 11.6$~\si{m} concert hall with a broadband reverberation time of \SI{1.46}{s}~\cite{Stade2012}. It was measured using a sequential 50-channel rigid-sphere microphone array with a radius of \SI{8.75}{cm}. \rev{The subspace decomposition used a block size of 64~samples (\SI{1.3}{ms}), a hop size of 8~samples and GSV averaging of 48~blocks. The other parameters were the same as previously.} Fig.~\ref{fig:caseStudy}~(b) shows the direct and residual subspace decomposition for the measurement that was performed using a PA loudspeaker located at the center of the stage.

The third SRIR was measured at the entrance of an anechoic chamber with the measurement loudspeaker placed within line-of-sight in the adjacent $6 \times 3.8 \times 2.8$~\si{m} office~\cite{McKenzie2021a}. The measurement again was performed using the Eigenmike em32 microphone array but in this case, the SRIR is provided as 25-channel SRIR in the SH domain. \rev{The subspace decomposition was performed in blocks of 32~samples (\SI{0.7}{ms}), with a hop size of 4~samples and the thresholds were calculated with a GSV averaging length of 32~blocks. The other parameters were the same as previously.} Fig.~\ref{fig:caseStudy}~(c) shows the direct and residual subspace decomposition for a measurement that was taken \SI{50}{cm} from the open door inside the anechoic chamber and contains strongly anisotropic reverberation.

In the case of the SRIRs from smaller rooms, cf.~Figs.~\ref{fig:caseStudy}~(a) and~(c), salient reflections are mainly extracted within the first \SI{30}{ms} but some weaker reflections are extracted until \SI{100}{ms} after the direct sound. In the case of the SRIR from the concert hall, Fig.~\ref{fig:caseStudy}~(b), salient reflections are extracted until \SI{200}{ms} after the direct sound. 
Although all three SRIRs, stemming from a small conference room, a concert hall, and from the transition between an office and an anechoic chamber, exhibit vastly different reverberation characteristics, the algorithm successfully separates the direct part and the residual in all three cases. Thus, the proposed algorithm proves to be applicable also when using measurement data from a variety of acoustic environments.

\section{Conclusion}\label{sec:conclusion}
In this work, we proposed a subspace method for the decomposition of SRIRs into a direct part, containing the direct sound and salient reflections, and a residual. The method does not rely on a specific microphone array geometry but the array configuration needs to guarantee a singular rank of the covariance matrix in the presence of a plane wave, which, for instance, is shown to be the case for spherical arrangements with a radius of \SI{4.2}{cm} and 24 or more microphones. \rev{The proposed method does not assume a specific wave model and does not rely on corresponding parameter estimates. It outperforms existing methods that rely on DOA estimation, beamforming, and the assumption of plane waves in all simulated scenarios, which include different rooms, microphone arrays, and ratios of direct sound to residual. It further generates lower direct part and residual errors than the compared methods in scenarios with two simultaneous reflections.} The proposed subspace decomposition can be applied to SH-domain SRIRs without modification and guarantees the perfect reconstruction of the original SRIR by summing up the direct part and the residual. \rev{The method facilitates novel ways of SRIR-based virtual acoustic rendering and might enhance the performance of established parameter estimation methods when applied as pre-processing. A reference implementation is provided at \texttt{https://github.com/thomasdeppisch/ SRIR-Subspace-Decomposition}.}

\section*{Acknowledgment}
Thomas Deppisch would like to thank Franz Zotter for numerous fruitful discussions about linear algebra.

\bibliographystyle{IEEEtran}
\bibliography{references,references_JA}

\begin{IEEEbiography}[{\includegraphics[width=1in,height=1.25in,clip,keepaspectratio]{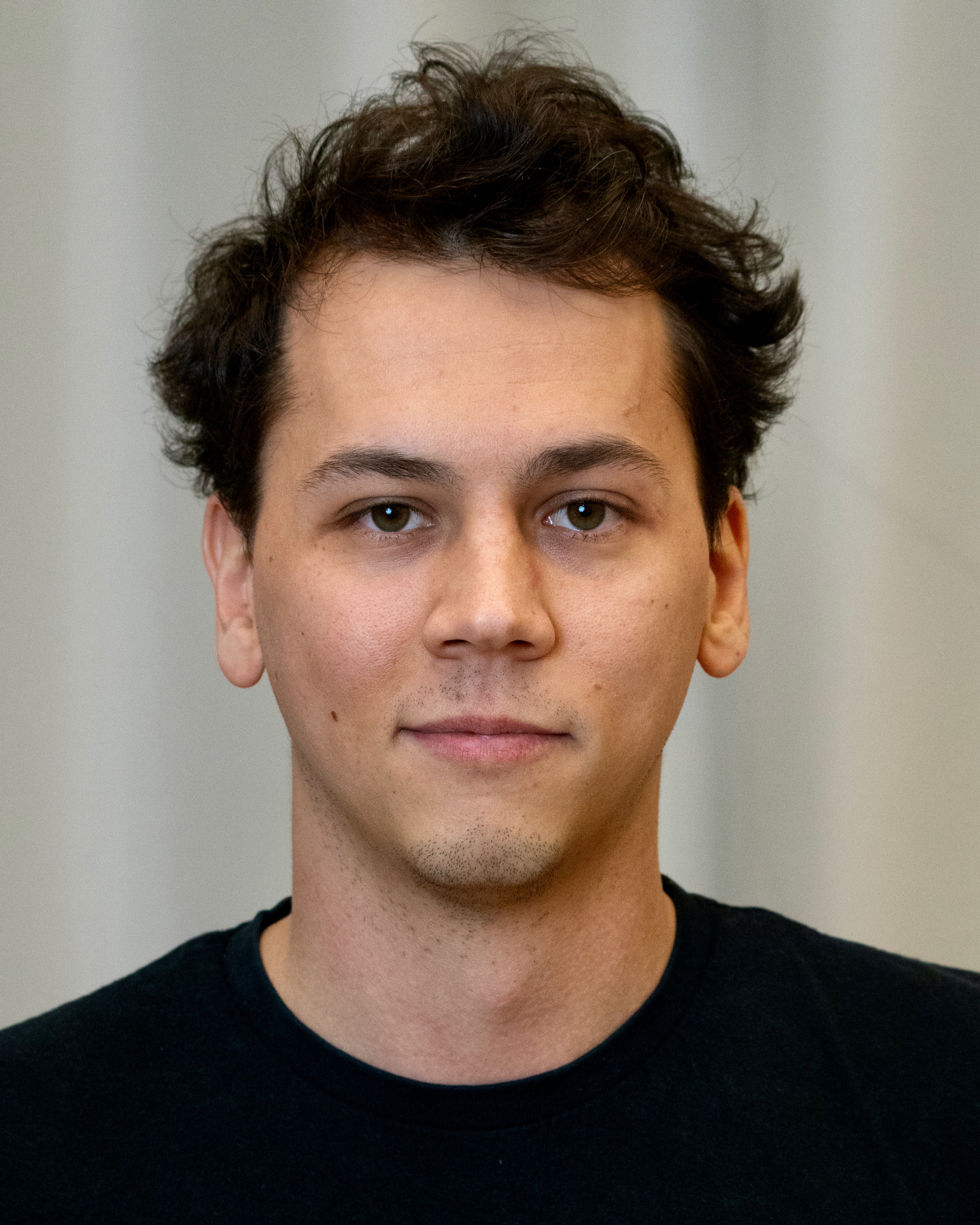}}]{Thomas Deppisch}
received the M.Sc. degree in Electrical Engineering and Audio Engineering jointly from Graz University of Technology and the University of Music and Performing Arts, Graz, Austria, in~2020. Since 2020, he has been working toward the Ph.D. degree at the Division of Applied Acoustics at Chalmers University of Technology, Gothenburg, Sweden. His main research interests lie in signal processing methods for the virtual reproduction of acoustic environments, including capture, analysis, rendering, and perception.
\end{IEEEbiography}
\begin{IEEEbiography}[{\includegraphics[width=1in,height=1.25in,clip,keepaspectratio]{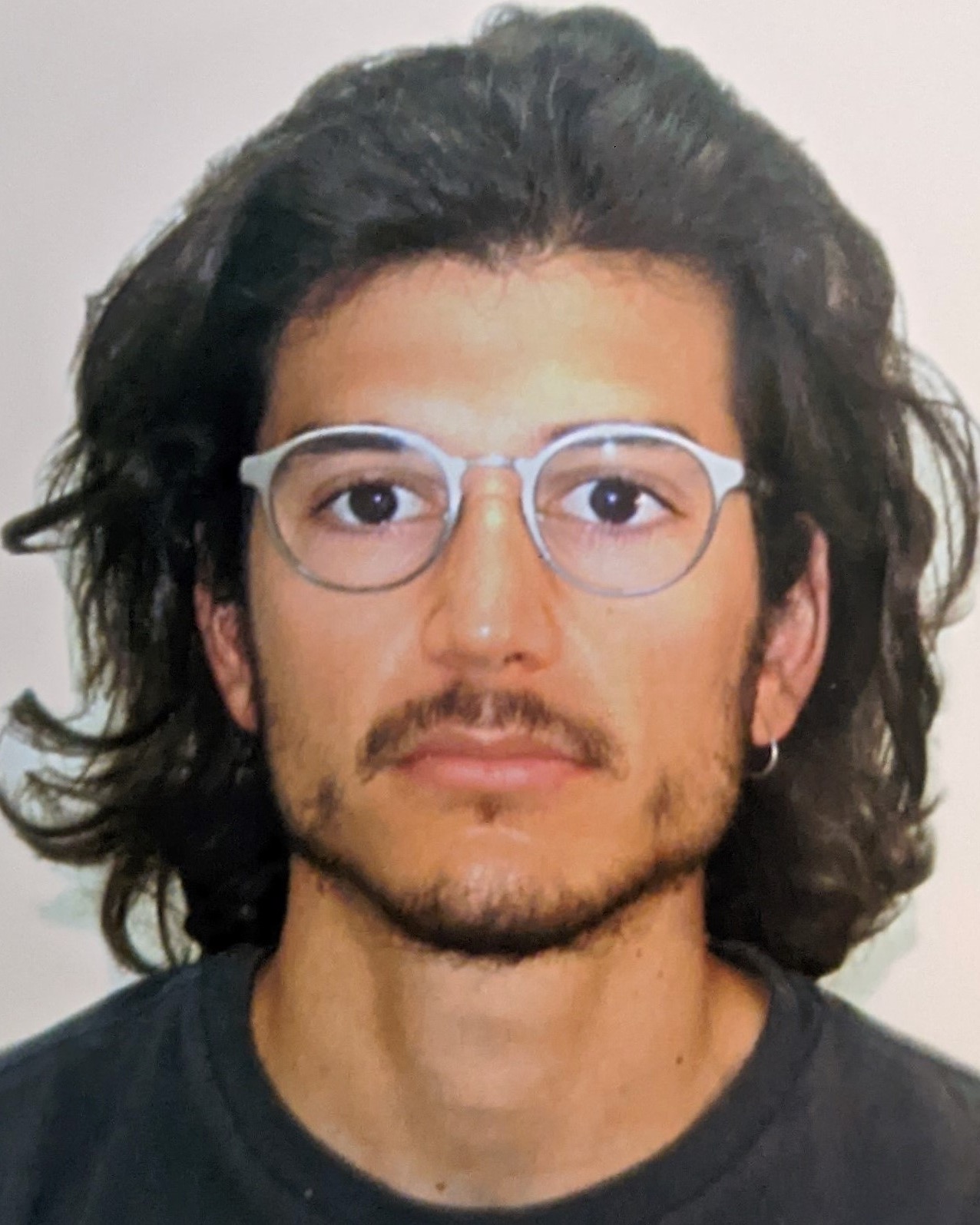}}]{Sebastià~V.~Amengual~Garí}
is currently a research scientist at Reality Labs Research working on room acoustics, spatial audio and auditory perception. He received a Diploma Degree in Telecommunications with major in Sound and Image in 2014 from the Polytechnic University of Catalonia (UPC) in 2014, completing his Master's Thesis at the Norwegian University of Science and Technology (NTNU). His doctoral work at the Detmold University of Music focused on investigating the interaction of room acoustics and live music performance using virtual acoustic environments. His research interests lie in the intersection of audio, perception and music.
\end{IEEEbiography}
\begin{IEEEbiography}[{\includegraphics[width=1in,height=1.25in,clip,keepaspectratio]{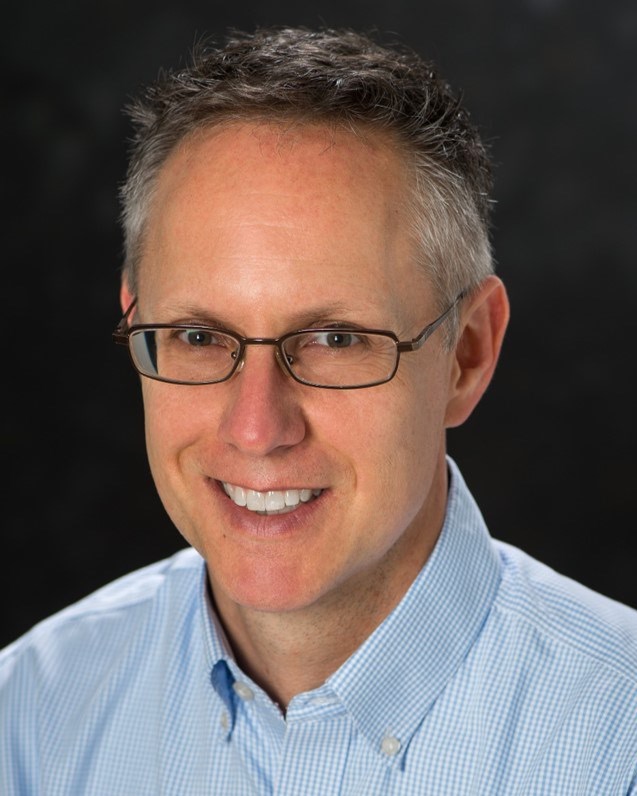}}]{Paul~Calamia}
is a research science manager on the Audio Team at Reality Labs Research, where he supports research in room acoustics for augmented-reality applications. Previously he was a member of the Technical Staff at MIT Lincoln Laboratory in the Bioengineering Systems and Technologies Group, with a focus on auditory health and hearing protection, and the Advanced Undersea Systems and Technology Group, working on sonar signal processing. His other prior positions include Assistant Professor in the Graduate Program in Architectural Acoustics at Rensselaer Polytechnic Institute in Troy, NY, and Consultant and Head of R\&D at Kirkegaard Associates in Chicago, IL. He holds a BS degree in mathematics from Duke University, an MS degree in electrical and computer engineering from the Engineering Acoustics Program at the University of Texas at Austin, and a Ph.D. in computer science from Princeton University.
\end{IEEEbiography}
\begin{IEEEbiography}[{\includegraphics[width=1in,height=1.25in,clip,keepaspectratio]{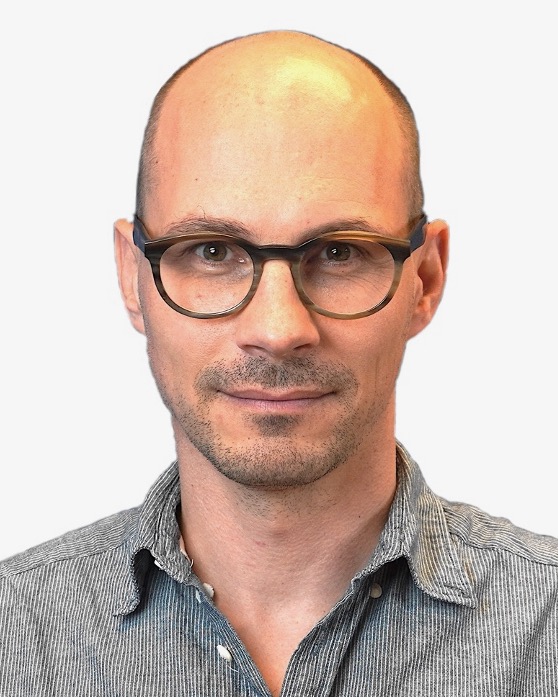}}]{Jens Ahrens}
has been an Associate Professor within the Division of Applied Acoustics at Chalmers since~2016. He received a Diplom (equivalent to a~M.Sc.) in Electrical Engineering and Audio Engineering jointly from Graz University of Technology and the University of Music and Performing Arts, Graz, Austria, in~2005. He completed his Doctoral Degree (Dr.-Ing.) at the Technische Universität Berlin,~Germany, in~2010.
From~2011 to~2013, he was a Postdoctoral Researcher at Microsoft Research in Redmond, Washington, USA, and in the fall and winter terms of~2015/16, he was a Visiting Scholar at the Center for Computer Research in Music and Acoustics~(CCRMA) at Stanford University, California,~USA.
He is an Associate Editor of the IEEE Signal Processing Letters and of the EURASIP Journal on Audio, Speech, and Music Processing.
\end{IEEEbiography}
\vfill

\end{document}